\documentclass[twocolumn]{aastex631}

\newif\ifedits
\editsfalse
\ifedits
    \renewcommand{\edit}[1]{{\color{red} #1}}
    
\else
    \renewcommand{\edit}[1]{#1}
    
\fi

\usepackage{booktabs}
\usepackage{array}
\usepackage{multirow}
\usepackage{float}
\usepackage{lipsum}

\newcolumntype{x}[1]{>{\centering\arraybackslash\hspace{0pt}}p{#1}}

\newcommand{\lc}[1]{\multicolumn{1}{|r}{#1}}
\newcommand{\lcTwo}[1]{\multicolumn{2}{|c}{#1}}

\usepackage{enumitem}

\begin{document}


\title{Metallicity Gradients in Modern Cosmological Simulations I: Tension Between Smooth Stellar Feedback Models and Observations}

\correspondingauthor{Alex M. Garcia} \\
\email{alexgarcia@virginia.edu}

\author[0000-0002-8111-9884]{Alex M. Garcia}
\affiliation{Department of Astronomy, University of Virginia,
530 McCormick Road, Charlottesville, VA 22904}
\affiliation{Virginia Institute for Theoretical Astronomy, University of Virginia, Charlottesville, VA 22904, USA}
\affiliation{The NSF-Simons AI Institute for Cosmic Origins, USA}

\author[0000-0002-5653-0786]{Paul Torrey}
\affiliation{Department of Astronomy, University of Virginia, 
530 McCormick Road, 
Charlottesville, VA 22904}
\affiliation{Virginia Institute for Theoretical Astronomy, University of Virginia, Charlottesville, VA 22904, USA}
\affiliation{The NSF-Simons AI Institute for Cosmic Origins, USA}

\author[0000-0003-0275-5506]{Aniket Bhagwat}
\affiliation{Max Planck Institut f\"ur Astrophysik, Karl Schwarzschild Stra{\ss}e 1, D-85741 Garching, Germany}

\author[0000-0002-1367-0949]{Ruby J. Wright}
\affiliation{International Centre for Radio Astronomy Research (ICRAR), M468, University of Western Australia, 35 Stirling Hwy, Crawley, WA 6009, Australia}

\author[0000-0002-4382-1090]{Qian-Hui Chen}
\affiliation{Research School of Astronomy \& Astrophysics, Australian National University, Canberra, Australia, 2611}
\affiliation{ARC Centre of Excellence for All Sky Astrophysics in 3 Dimensions (ASTRO 3D), Australia}

\author[0000-0002-3247-5321]{Kathryn Grasha}
\affiliation{Research School of Astronomy \& Astrophysics, Australian National University, Canberra, Australia, 2611}
\affiliation{ARC Centre of Excellence for All Sky Astrophysics in 3 Dimensions (ASTRO 3D), Australia}
\affiliation{Visiting Fellow, Harvard-Smithsonian Center for Astrophysics, 60 Garden Street, Cambridge, MA 02138, USA}

\author[0000-0003-3569-4092]{Sophia Ridolfo}
\affiliation{ARC Centre of Excellence for All Sky Astrophysics in 3 Dimensions (ASTRO 3D), Australia}
\affiliation{Harvard-Smithsonian Center for Astrophysics, 60 Garden Street, Cambridge, MA 02138, USA}

\author[0000-0002-0799-0225]{Z. S. Hemler}
\affiliation{Department of Astrophysical Sciences, Princeton University,  Peyton Hall,  Princeton, NJ, 08544, USA }

\author[0000-0002-5222-1337]{Arnab Sarkar}
\affiliation{Kavli Institute for Astrophysics and Space Research, Massachusetts Institute of Technology, 70 Vassar Street, Cambridge, MA 02139, USA}

\author[0000-0002-4469-2518]{Priyanka Chakraborty}
\affiliation{Harvard-Smithsonian Center for Astrophysics, 60 Garden Street, Cambridge, MA 02138, USA}

\author[0000-0002-7524-374X]{Erica J. Nelson}
\affiliation{Department for Astrophysical and Planetary Science, University of Colorado, Boulder, CO 80309, USA}

\author[0000-0003-4792-9119]{Ryan L. Sanders}
\affiliation{Department of Physics and Astronomy, University of Kentucky, 505 Rose Street, Lexington, KY 40506, USA}

\author[0000-0002-6748-2900]{Tiago Costa}
\affiliation{School of Mathematics, Statistics and Physics, Newcastle University, Newcastle upon Tyne, NE1 7RU, UK}

\author[0000-0001-8593-7692]{Mark Vogelsberger}
\affiliation{Department of Physics and Kavli Institute for Astrophysics and Space Research, Massachusetts Institute of Technology, Cambridge, MA 02139, USA}

\author[0000-0001-8152-3943]{Lisa J. Kewley}
\affiliation{Harvard-Smithsonian Center for Astrophysics, 60 Garden Street, Cambridge, MA 02138, USA}

\author[0000-0002-1768-1899]{Sara L. Ellison}
\affiliation{Department of Physics \& Astronomy, University of Victoria, Finnerty Road, Victoria, British Columbia, V8P 1A1, Canada}

\author[0000-0001-6950-1629]{Lars Hernquist}
\affiliation{Harvard-Smithsonian Center for Astrophysics, 60 Garden Street, Cambridge, MA 02138, USA}



\begin{abstract}
The metallicity of galaxies, and its variation with galactocentric radius, provides key insights into the formation histories of galaxies and the physical processes driving their evolution.
In this work, we analyze the radial metallicity gradients of star forming galaxies in the EAGLE, Illustris, IllustrisTNG, and SIMBA cosmological simulations across a broad mass ($10^{8.0}M_\odot\leq M_\star \lesssim10^{12.0}M_\odot$) and redshift ($0\leq z\leq8$) range.
We find that all simulations predict strong negative (i.e., radially decreasing) metallicity gradients at early cosmic times, likely due to their similar treatments of relatively smooth stellar feedback \edit{not providing sufficient mixing to quickly flatten gradients}.
The strongest redshift evolution occurs in galaxies with stellar masses of $10^{10.0}-10^{11.0}M_\odot$, while galaxies with stellar masses $< 10^{10}M_\odot$ and $>10^{11}M_\odot$ exhibit weaker redshift evolution.
Our results of negative gradients at high-redshift contrast with the many positive and flat gradients in the $1<z<4$ observational literature.
At $z>6$, the negative gradients observed with JWST and ALMA are flatter than those in simulations, albeit with closer agreement than at lower redshift.
Overall, we suggest that these smooth stellar feedback galaxy simulations may not sufficiently mix their metal content \edit{radially}, and that either stronger stellar feedback or additional subgrid turbulent metal diffusion models may be required to better reproduce observed metallicity gradients.
\end{abstract}



\section{Introduction} \label{sec:intro}

The baryon cycle that occurs within galaxies is what drives their evolution \citep[e.g.,][]{Somerville_Dave_2015,Tumlinson_2017,Ayromlou_2023,Wright_2024}.
Gas is accreted onto galaxies from the circumgalactic medium (CGM, \citeauthor{Koeppen_1994} \citeyear{Koeppen_1994}; \citeauthor{Keres_2005} \citeyear{Keres_2005}), cools in the interstellar medium (ISM) to form stars \citep{McKee_2007,Kennicutt_2012}, and follows galactic winds and outflows generated by feedback from those stars \citep{Muratov_2015,Veilleux_2020} to either escape the system entirely or begin the process anew by raining back down as recycled gas.
It is not surprising, then, that perhaps the most sensitive tracer of both the local and global processes in galaxies is that very same gas content.
The gas-phase metallicity, in particular, is indispensable for decoding the underlying evolutionary processes and physical conditions within galaxies \citep[see reviews by][]{Kewley_2019,Maiolino_Mannucci_2019}.

A galaxy's metal content is oftentimes not homogeneously distributed.
Galaxies in the local Universe typically have more metals in their nuclear regions than the outskirts \citep[e.g.,][]{Searle_1971,Zaritsky_1994,Kewley_2010,Berg_2015,Ho_2015,Belfiore_2017,Grasha_2022,Chen_2023,Khoram_2025}.
This is most likely a product of galaxies forming inside-out: the stellar populations in the interior form and evolve earlier than those of the outskirts, chemically enriching the inner regions first \citep[][etc]{Prantzos_2000,Pilkington_2012,Perez_2013,Tissera_2019}.\ignorespaces
\footnote{\ignorespaces
~\!\edit{Since much of the enrichment from massive stars occurs rapidly ($\lesssim100$ Myr), it is not necessarily the inside-out growth that leads to persistently negative gradients.
Rather the ratio of metal enrichment to metal-diluting gas across radii is what sets (and changes) a metallicity gradient \citep[see, e.g.,][]{Garcia_2023, Graf_2024}.
}}
This radially decreasing distribution of metals is commonly parameterized by a single linear fit\ignorespaces
\footnote{\ignorespaces
We note that there have also been efforts -- in both observations and simulations -- to fit metallicity profiles with multiple regressions \citep[][]{Tapia_2022,Chen_2023,Chen_2024,Tapia_Contreras_2025}, splines \citep{Garcia_2023}, and/or non-parametric methods \citep[see, e.g.,][]{Kirby_2011,Acharyya_2024}.
}
(in logarithmic metallicity) and thus characterized by its gradient -- the slope of that regression.
Given the simplistic inside-out evolutionary view, galaxies should have ``negative'' gradients.
Typical values for the negative gradients of ``normal'' observed galaxies in the local Universe are around $-0.05\pm0.05~{\rm dex/kpc}$ \citep[e.g.,][]{Rupke_2010b,Sanchez_2014,Sanchez_Menguiano_2016,Grasha_2022}.
Not all galaxies exhibit negative metallicity gradients, however; galaxies -- both in simulations and observations -- undergoing strong inflow/outflow events, pristine gas inflows, and/or mergers can exhibit flattened or even positive gradients \cite[see, e.g.,][]{Rupke_2010a,Rupke_2010b,Torrey_2012,Ceverino_2016,Tissera_2022,Venturi_2024}. 
Moreover, observations with limited angular resolution, signal-to-noise, and/or spectral resolution add complexity and seem to artificially flatten gradients \citep[see work by][]{Yuan_2013,Mast_2014,Poetrodjojo_2019,Acharyya_2020,Grasha_2022}.
Galaxies of different masses also exhibit different gradients.
\cite{Belfiore_2017}, using Sloan Digital Sky Survey (SDSS) IV, suggest that galaxies with lower masses $(M_\star \sim10^{9.5}M_\odot)$ and higher masses $(M_\star \sim10^{11.0}M_\odot)$ have flatter gradients than systems of intermediate mass $(M_\star \sim10^{10.5}M_\odot)$, which have the strongest negative gradients.

The picture becomes less clear at higher redshift.
Early studies of galaxies around cosmic noon ($z=1-3$) reported a wide variety of positive, flat, and/or negative gradients \citep[e.g.,][with gradients typically falling between $\pm0.1~{\rm dex/kpc}$]{Cresci_2010,Yuan_2011,Queyrel_2012,Swinbank_2012,Troncoso_2014,Wuyts_2016,Wang_2017,Curti_2020,Simons_2021,Gillman_2022,Li_2022, Dutta_2024}.
Here, the limited spatial and spectral resolution becomes even more of a hindrance without the use of gravitationally lensed systems \citep[as in][]{Jones_2010,Jones_2013,Yuan_2011} or adaptive optics \citep[as in][]{Swinbank_2012}.
It was therefore extremely difficult, up until recently, to make any robust statements about metallicity gradients during this epoch.
As such, no clear consensus on the metallicity gradient evolution of galaxies in the Universe has been established.

The recent launch and successful deployment of JWST serves to provide unprecedented fidelity at higher redshifts to help alleviate the systematic issues with the previous generation of observations.
Some work has already been done on characterizing gradients with JWST at $z=1-4$ \citep{Wang_2022,Ju_2024,Morishita_2024,Rodriguez_Del_Pino_2024} and beyond \citep[$z>6$;][]{Arribas_2024,Venturi_2024}.
While the sample sizes remain limited, the tentative scenario is that, indeed, there are a wide assortment of metallicity gradients in the early Universe.
However, unlike those of previous $z>3$ studies, not all observed metallicity gradients are positive.
In fact, to date, only one system with a positive gradient has been observed at $z>6$ \citep[][though a few have uncertainties that encompass positive gradient values]{Venturi_2024}.

The current theoretical understanding of the evolution of gradients from simulations \citep[][etc]{Gibson_2013,Taylor_Kobayashi_2017, Tissera_2016,Tissera_2019,Tissera_2022,Hemler_2021,Acharyya_2024,Ibrahim_2025} and other analytic models \citep[e.g.,][]{Mott_2013,Kubryk_2015,Molla_2019,Sharda_2021a,Sharda_2021b} is similarly heterogeneous.
Some models predict that gradients should become positive at $z>3$ \citep[][though this depends on assumptions about star formation efficiencies]{Mott_2013}, some predict that gradients should become \edit{flatter with decreasing redshift} \citep[][though the detailed evolution depends strongly on stellar feedback implementation of the model]{Gibson_2013,Taylor_Kobayashi_2017,Hemler_2021}, some predict that \edit{gradients should get {\it steeper} (more negative) closer to $z=0$} \citep[][]{Ma_2017,Bellardini_2021,Bellardini_2022,Sharda_2021b,Sun_2024,Graf_2025}, \edit{and others still predict virtually no redshift evolution \citep{Tissera_2019}}.
Moreover, relatively little work has been done analyzing metallicity gradients at very high redshift ($z\gtrsim3$; although it is not completely unexplored, see, e.g., \citeauthor{Taylor_Kobayashi_2017} \citeyear{Taylor_Kobayashi_2017}; \citeauthor{Ibrahim_2025} \citeyear{Ibrahim_2025}).
How metallicity gradients evolve through time is thus very much an open question.

Beyond the lack of theoretical predictions at high redshift, there exists a subtle, yet troubling, tension between two similar models in simulations even at $z\leq3$.
Briefly, the current simulation landscape contains two different approaches:
(i) high resolution models with an explicit treatment of the ISM that gives rise to strong stellar feedback and
(ii) lower resolution simulations that treat the ISM with a `subgrid' prescription that give rise to weaker stellar feedback (see \citeauthor{Vogelsberger_2020} \citeyear{Vogelsberger_2020} and Section~\ref{subsec:simulations} for a more complete discussion).
The work presented in \cite{Gibson_2013} shows that models with stronger feedback should see less metallicity gradient evolution (i.e., flatter gradients at high-redshift) than those with the weaker feedback.
For example, \citeauthor{Hemler_2021} (\citeyear{Hemler_2021}; see their Figure 6) present gradients from the IllustrisTNG subgrid ISM model, finding that negative gradients steepen further back in time (at $z\leq3$).
\cite{Hemler_2021} cite the relatively smooth stellar feedback in IllustrisTNG for the assembly of strong negative gradients at early times allowing \edit{for negative gradients to persist}.
On the other hand, strong feedback models, like that of FIRE (Feedback In Realistic Environments; \citeauthor{Hopkins_2014} \citeyear{Hopkins_2014}, \citeyear{Hopkins_2018}, \citeyear{Hopkins_2023}) find \edit{comparatively} little redshift evolution of gradients owing to the large episodic feedback events \citep[][\edit{in more detail, these authors find that gradients get slightly steeper at lower redshift}]{Ma_2017,Sun_2024,Graf_2025}, in line with the strong feedback model of \cite{Gibson_2013}.
However, \cite{Tissera_2022}, using the EAGLE model (which is of the weaker feedback, subgrid ISM variety), find that there is virtually no redshift evolution of metallicity gradients to $z\sim2.5$ (see Figure 2 in \citeauthor{Tissera_2022} \citeyear{Tissera_2022}).
This EAGLE result is in direct tension with both the \cite{Gibson_2013} and \cite{Hemler_2021} predictions since EAGLE has this weaker feedback model.
The tension between EAGLE and IllustrisTNG gradient evolution is yet unexplored but potentially presents a huge uncertainty in our understanding of what drives the chemical evolution of galaxies.
It is therefore critical to establish some notion of the chemical enrichment of galaxies from simulation models to rectify our understanding of galaxy evolution as well as constrain the nature of feedback implemented in future simulation efforts.

In this paper, we begin this effort to provide a more complete and comprehensive view of the spatial variations of the metal content within current galaxy simulation paradigm.
This work, which is the first of a series, focuses on several of the current class of large box simulations that contain a large dynamic range of galaxies.
The remainder of this paper is organized as follows:
in Section~\ref{sec:methods} we describe each simulation model, our selection criteria, and methodology for deriving metallicity gradients.
In Section~\ref{sec:results}, we present the redshift evolution of the metallicity gradients in each simulation model as well as break down evolution in different stellar mass bins.
In Section~\ref{sec:discussion}, we suggest a possible resolution to the \cite{Hemler_2021}-\cite{Tissera_2022} literature tension and compare our results to recent observations.
Finally, in Section~\ref{sec:conclusion}, we state our conclusions.

\section{Methods}\label{sec:methods}

\subsection{Simulations}\label{subsec:simulations}

This work uses data products from the EAGLE, Illustris, IllustrisTNG, and SIMBA cosmological simulations of galaxy formation and evolution.
Each of these simulations models a range of astrophysical processes including gravity, cosmology, star formation, stellar evolution, stellar feedback, chemical enrichment of the ISM, and black hole growth and feedback.
We therefore dedicate this section to describing each model, cautioning that this section is not meant to be a complete enumeration of all of the details of each model; rather it is a brief description of each of the physical models with details pertinent to the results presented in this work.

We first note that these models share the commonality of having a ``subgrid'' prescription for the ISM, owing to their limited resolution.
This subgrid prescription, modeled with an effective equation of state, sets the behavior of the cold, dense gas where star formation takes place.
As a consequence, all the models in this work have relatively `smooth' stellar feedback that is persistent in time, yet not terribly destructive.
This allows for relatively smooth gas cycling as galaxies assemble \citep[see, e.g.,][though there are differences in the implementations, see \citeauthor{Wright_2024} \citeyear{Wright_2024} for a careful examination]{Torrey_2018,Garcia_2024a,Garcia_2024b,Garcia_2025}.
This is contrasted with high-resolution explicit ISM models that directly model the sites of star formation and feedback, such as the FIRE simulations \citep{Hopkins_2014,Hopkins_2018,Hopkins_2023}.
These explicit ISM models naturally produce much stronger feedback resulting in episodic bursts that eject gas far into the CGM around the galaxy \citep{Muratov_2015,Muratov_2017,Angles_Alcazar_2017b,Pandya_2021}.
We discuss the potential implications of using this smooth feedback paradigm in Section~\ref{subsubsec:3<z<4}.

Moreover, none of \edit{the} simulations \edit{analyzed in this work} have a method by which unresolved turbulence can exchange mass and metals throughout the galaxy.
Mass can move along with bulk galactic winds, but the small-scale turbulent eddies that drive diffusion within the ISM \citep{Smagorinsky_1963,Shen_2010,Semenov_2016,Su_2017,Escala_2018,Semenov_2024} are not modeled in any of the simulations in this work.
We consider what role this (lack of) unresolved turbulence may play in setting the gradients more carefully in Section~\ref{subsubsec:z>6}.

\subsubsection{EAGLE}
\label{subsubsec:eagle}

The EAGLE \citep{Crain_2015, Schaye_2015, McAlpine_2016} simulations are built upon a modified version of the smooth particle hydrodynamics (SPH) code {\sc gadget-3} \citep{Springel_2005} code called {\sc anarchy} \citep{Schaye_2015}.
As mentioned above, the ISM in EAGLE is treated with an effective equation of state (\citeauthor{Schaye_DallaVechhia_2008} \citeyear{Schaye_DallaVechhia_2008}).
Star formation in EAGLE is thus restricted to gas cells with
\begin{equation}
    \label{eqn:EAGLE_SF}
    n_{\rm H} \geq 10^{-1}\,\edit{{\rm cm}^{-3}}~\left(\frac{Z}{0.002}\right)^{-0.64}
\end{equation}
and $\log T < \log T_{\rm eos} + 0.5$, where $n_{\rm H}$ is the hydrogen number density of the gas, $T$ is the gas temperature, $Z$ is the metallicity of the gas \edit{(i.e., the fraction of gas mass in elements heavier than Helium)}, and $T_{\rm eos} = 8\times10^3~{\rm K}$ \citep{Schaye_2004,Dalla_Vecchia_2012,Schaye_2015}.
Stars form from this gas according to a \cite{Chabrier_2003} initial mass function (IMF) and evolve according to \cite{Wiersma_2009b} evolutionary tracks.
Mass and metals are sent back into the ISM via asymptotic giant branch (AGB) winds as well as Type Ia and II supernovae.
The EAGLE model explicitly tracks the production and evolution of eleven chemical species (H, He, C, N, O, Ne, Mg, Si, Fe, S, and Ca) coming from yield tables in \cite{Marigo_2001} for AGB winds, \cite{Thielemann_2003} for Type Ia supernovae, and \cite{Portinari_1998} for Type II supernovae.
The fluid is modeled as discrete particles in SPH codes (like that of EAGLE) and as such newly ejected metals are locked into the cell in which they form.

In more detail, the feedback from stars in EAGLE is modeled through stochastic thermal energy \citep{Kay_2003,Dalla_Vecchia_2012}.
Thermal energy is injected locally to the $N$ nearest particles assuming some amount of energy, $f_{\rm th}$, is released unit mass formed ($f_{\rm th}$ is set assuming [i] a \citeauthor{Chabrier_2003} \citeyear{Chabrier_2003} IMF, [ii] the energy of a single supernova is $10^{51}~{\rm erg}$, and [iii] that stars with mass $>6M_\odot$ explode; \citeauthor{Dalla_Vecchia_2012} \citeyear{Dalla_Vecchia_2012}).
Importantly, this thermal energy injection is not a constant through all time, nor all densities in the ISM.
The total thermal energy released increases with star particle birth cloud density and decreases with metallicity.
This behavior is asymptotic taking a maximum of $3\times f_{\rm th}$ at high density (and/or low metallicity) and $0.3\times f_{\rm th}$ at low density (and/or high metallicity).
Since galaxies at earlier times tend to be at lower metallicity, the effect is that stellar feedback is more efficient at earlier times in EAGLE.

The full EAGLE suite is comprised of several different runs with varied resolution, box size, and calibrations.
In this work, we analyze the flagship $(67.8~{\rm Mpc}/h)^3$ high-resolution box ({\sc RefL0100N1504}; hereafter synonymous with EAGLE itself), which has $2\times1504^3$ particles and an initial baryon mass resolution of $1.81\times10^6M_\odot$.

\subsubsection{Illustris}

The original Illustris suite of simulations \citep{Vogelsberger_2013,Vogelsberger_2014a,Vogelsberger_2014b,Genel_2014,Torrey_2014} was run using the Moving Voronoi Mesh (MVM) code {\sc arepo} \citep{Springel_2010}.
The dense, star forming ISM is treated with the \cite{Springel_Hernquist_2003} effective equation of state in Illustris.
Stars form stochastically in the dense ($n_{\rm H}>0.13~{\rm cm}^{-3}$) ISM according to a \cite{Chabrier_2003} IMF.
The stellar evolutionary tracks are taken from \cite{Portinari_1998} and depend on both the mass and metallicity of the stars.
The Illustris model explicitly tracks nine chemical species (H, He, C, N, O, Ne, Mg, Si, and Fe).
Stars return their mass and metals to the ISM through AGB winds and supernovae.
The metal yields for AGB winds come from \cite{Karakas_2010}, Type Ia supernovae from \cite{Thielemann_2003}, and Type II supernovae from \cite{Portinari_1998}.
Finally, we note that the Voronoi mesh structure of {\sc arepo} naturally allows for metals to exchange across boundaries as the cells deform and move.
However, as mentioned previously, there is no model for allowing metals to do so from unresolved turbulence.

Winds produced by stellar feedback in Illustris are decoupled from hydrodynamics \citep{Vogelsberger_2013}.
A star-forming gas cell from the region producing the feedback is selected probabilistically to transform into a ``wind particle''.
This wind particle then travels (with gravitational interactions, but without hydrodynamic forces) until it reaches a certain density threshold or a fixed time has elapsed.
The wind particle is then dissolved, dumping its mass, momentum, thermal energy, and metals into the gas cell where it ends up.
Wind particles are launched in a bipolar manner, with the direction given by $\vec{v}\times\nabla\Phi$ (where $\vec{v}$ and $\nabla\Phi$ are the velocity and acceleration, respectively, of the original gas cell in the rest frame of the halo) and the velocity based on the local 1D dark matter velocity dispersion.
The hydrodynamically decoupled winds approach is fundamentally different to that of EAGLE. 
Whereas the stochastic thermal implementation of EAGLE imparts energy locally into neighboring particles, the hydrodynamically decoupled wind particles of Illustris are non-local by construction.
Moreover, this implementation directly ties the wind rate to the star formation rate of the gas.

The full Illustris suite is comprised of several $(75~{\rm Mpc}/h)^3$ boxes of varying resolution.
Here we analyze the highest resolution run (Illustris-1, hereafter synonymous with Illustris itself) which has $2\times1820^3$ particles and an initial baryon mass resolution of $1.26\times10^6M_\odot$.

\subsubsection{IllustrisTNG}
\label{subsubsec:TNG}

IllustrisTNG \citep[hereafter TNG;][]{Marinacci_2018,Naiman_2018,Nelson_2018,Pillepich_2018b,Springel_2018,Pillepich_2019, Nelson_2019a, Nelson_2019b} is a suite of cosmological box simulations and is the successor to the original Illustris simulations.
The two models are thus very similar in spirit and there are key similarities, as well as differences, between the models that we will enumerate in this section \citep[see][for a detailed comparison of the two models]{Weinberger_2017,Pillepich_2018a}.
Star formation in TNG follows from the same \cite{Springel_Hernquist_2003} effective equation of state as Illustris.
Moreover, the stars formed follow the same \cite{Chabrier_2003} IMF.
There are changes in the treatment of stellar feedback as well as chemical enrichment, however.
The Illustris model set the minimum mass for core-collapse supernovae at $6M_\odot$, whereas TNG raises this to $8M_\odot$.
This accounts for a 30\% decrease in Type II supernovae in TNG.
TNG tracks the same nine chemical species as Illustris, with an additional tenth ``other metals'' field to account for untracked metals.
Many of the metal yield tables are updated from Illustris.
TNG adopts yields for Type Ia supernovae from \cite{Nomoto_1997}, Type II supernovae from \cite{Portinari_1998} and \cite{Kobayashi_2006}, and AGB winds from \cite{Karakas_2010}, \cite{Doherty_2014}, and \cite{Fishlock_2014}.
Just as in Illustris, TNG has no built-in model for the unresolved turbulence in the ISM beyond that naturally present in its MVM implementation.

As TNG is a similar model to Illustris, stellar feedback is implemented in largely the same manner as Illustis (see previous section).
There two major updates in the TNG model, however.
First, the winds of TNG are launched isotopically -- opposed to the bipolar winds of Illustris.
Second, the TNG model introduces both a minimum wind speed and a redshift-scaling to the winds.
The redshift-scaling winds help suppress low-redshift star formation while the minimum wind speed increasing the efficacy of feedback at high redshifts.

The full suite of TNG simulations is comprised of several different resolution runs as well as varied box sizes.
We use the highest resolution run of the $(35~{\rm Mpc}/h)^3$ volume (i.e., TNG50-1), which has $2\times2160^3$ resolution elements and an initial baryon mass resolution of $8.5\times10^4M_\odot$.
We note that we will use the term TNG synonymously with the TNG50-1 simulation throughout this work.
Additionally, we show a comparison to other box sizes and resolutions from the TNG suite in Appendix~\ref{appendix:resolution}.

\subsubsection{SIMBA}

The SIMBA simulations \citep{Dave_2019} are the successor to the MUFASA simulations \citep{Dave_2016} and are run using the meshless finite mass (MFM) code {\sc gizmo} \citep{Hopkins_2015}.
Star formation in SIMBA is set by a molecular column density (and metallicity) of the gas cells with densities $n_{\rm H}\geq0.13~{\rm cm}^{-3}$ \citep*[adapted from][]{Krumholz_2009,Krumholz_2011}.
Newly formed star particles inherit their metallicity from their natal gas.
SIMBA explicitly tracks the same eleven chemical species as EAGLE.
Feedback from stars is implemented in the form of AGB winds consistent with a \cite{Chabrier_2003} IMF as well as Type Ia and II supernovae.
The metal yield tables come from \cite{Oppenheimer_Dave_2006} for AGB winds, \cite{Iwamoto_1999} for Type Ia supernovae, and \cite{Nomoto_2006} for Type II supernovae.
Unique to the SIMBA model (among simulations in this work) is that it also includes an explicit tracking of dust production, growth, and destruction (following \citeauthor{Dwek_1998} \citeyear{Dwek_1998}).
This is important for the purposes of this work since some of the metals produced in stars will be locked in dust and therefore not be accounted for in the gas-phase metallicity of the system.
MFM codes, such as {\sc gizmo}, also do {\it not} have a natural scheme for advecting metals from particle to particle.

The hydrodynamically decoupled stellar feedback in SIMBA follows more closely to that of Illustris and TNG than EAGLE; however, there are a few key additions in the SIMBA model.
Wind particles are randomly assigned as `cool' ($T\approx10^3~{\rm K}$) or `hot' (temperature set by the supernova energy minus the wind particle's kinetic energy) such that $30\%$ of the wind particles are in the `hot' phase.
Also unique to SIMBA is that the mass loading factors of the winds are adopted using the results of higher resolution models that more explicitly resolve the behavior of winds in the ISM \citep{Angles_Alcazar_2017b}.
At high redshift ($z>3$), however, these mass loading factors are manually suppressed to allow for the growth of galaxies.
Another novel addition in the SIMBA model is metal-loaded winds.
Wind particles in SIMBA extract metals from nearby particles as they move.

The full SIMBA suite consists of several different box size runs.
In this work, we make use of a $(50~{\rm Mpc}/h)^3$ box with $2\times512^3$ particles (m50n512; hereafter simply SIMBA) and an initial baryon mass resolution of $1.28\times10^7M_\odot$.

\subsection{Selection Criteria}\label{subsec:selection_criteria}

We utilize {\sc subfind} \citep{Springel_2001} catalogs for each simulation to identify gravitationally bound substructure.
We apply a resolution cut of $\sim\!10^3$ to both the gas and stellar elements for a galaxy to be considered ``well-resolved''.
In TNG, this corresponds to $M_{\star} ~({\rm and}~M_{\rm gas}) > 10^8 M_\odot$ whereas in EAGLE and Illustris it is $M_\star > 10^{9} M_\odot$ and in SIMBA it is $M_\star > 10^{10} M_\odot$.
We discuss the implications that this selection criterion has for the results of this work in Section~\ref{subsection:evolution_by_mass} as well as the dependence on simulation resolution in Appendix~\ref{appendix:resolution}.

We further require that each galaxy has a ${\rm SFR}>0~[M_\odot/{\rm yr}]$.
We note that our gradient definitions require that the SFR be significant and spatially extended (see discussion below in Section~\ref{subsec:gradient_definitions}).
This tends to naturally raise the minimum required ${\rm SFR}$.
Regardless, we do not expect slight changes to this star forming galaxy selection (e.g., a specific SFR cut or star forming main sequence cut as in other works) to significantly change the samples which we draw.
Rather, the ${\rm SFR}>0~[M_\odot/{\rm yr}]$ requirement is a practical cut made such that: (i) the galaxy would likely contain bright emission lines used in observational surveys and (ii) our definitions of the metallicity gradient region are well-posed.

We also note that we restrict the sample to only central galaxies -- i.e., the most massive in a ``group''.

\edit{
Finally, we note that we are {\it not} explicitly tracking the evolution of individual galaxies through time in the simulations.
Indeed, individual galaxies may come into and out of our sample as they lose gas content, transition from star forming to quiescent, and/or merge into larger systems are no longer the central in their halo
Rather, this work compares the redshift evolution of gas-phase metallicity gradients of the populations of galaxies in each simulation.
}

\subsection{Metallicity Gradient Definitions}\label{subsec:gradient_definitions}

\begin{figure}
    \centering
    \includegraphics[width=\linewidth]{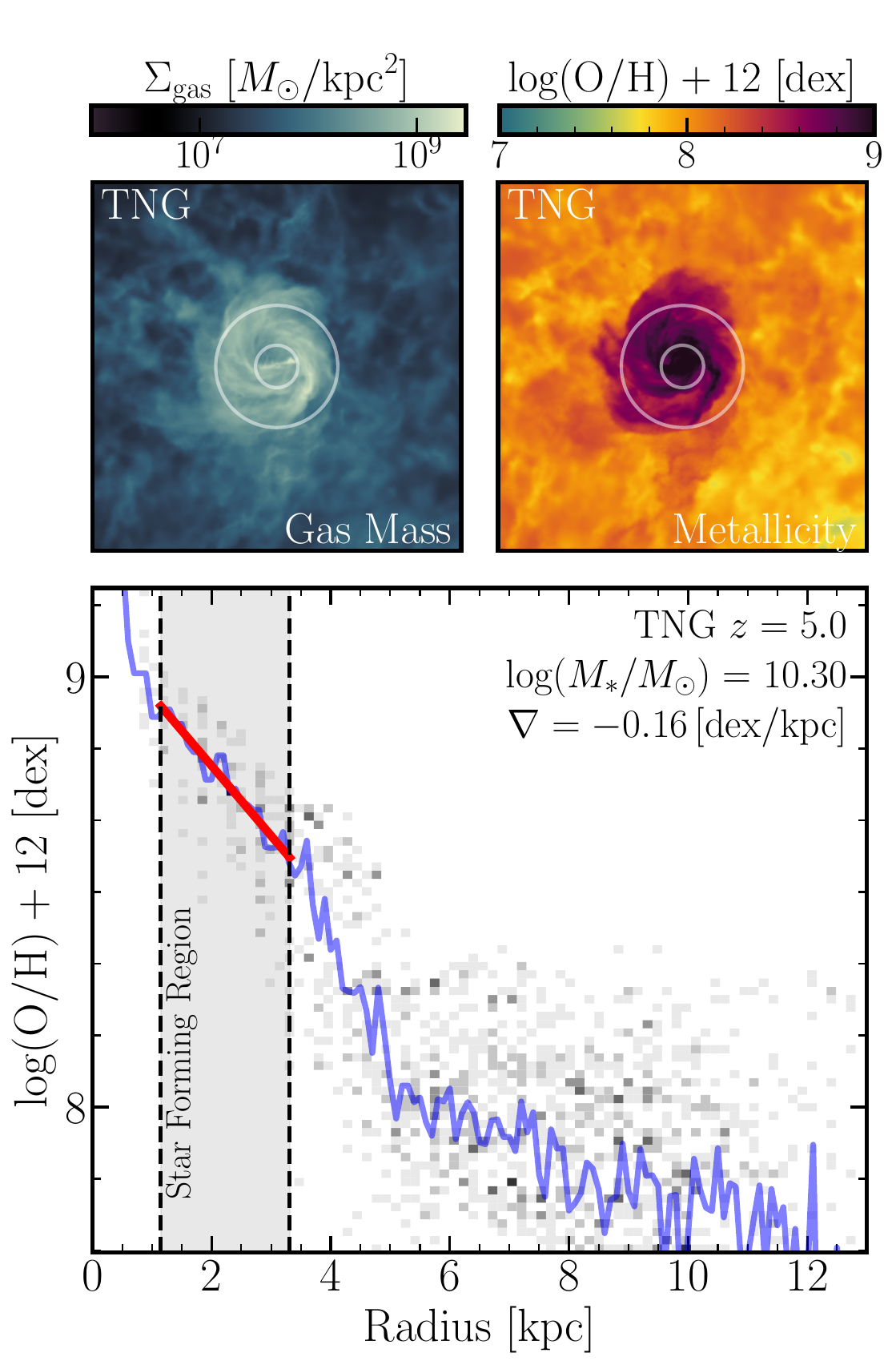}
    \caption{{\bf Demonstration of Metallicity Gradient Derivation.}
    Projections of a galaxy at $z=5$ from TNG in gas mass (top left) and metallicity (top right) as well as the resulting radial profile after creating $0.5~{\rm kpc}\times0.5~{\rm kpc}$ pixel maps of the system (gray 2D histogram in bottom panel).
    The median 1D metallicity profile is shown in the blue line in the bottom panel.
    The gray shaded region represents the ``star forming region'' (see Section~\ref{subsec:gradient_definitions} for definition) and is the region over which we fit the linear regression to obtain the metallicity gradient, $\nabla$ (note that the circles in the top panels also show this star forming region, roughly corresponding to the disk of the galaxy).
    The straight red line within the star forming region corresponds to the best-fit gradient (in this case $\nabla=-0.16$ dex/kpc).
    }
    \label{fig:methods_fig}
\end{figure}

The methodology related to defining metallicity gradients in the simulations derives heavily from a combination of previous theoretical studies by \cite{Ma_2017} and \cite{Hemler_2021}.
We utilize these well-tested methods in order to make as fair a comparison to observational studies as possible.
We provide a visual summary of the definitions for a TNG galaxy at $z=5$ in Figure~\ref{fig:methods_fig}.

We first center the galaxies by placing the location of the particle with the potential minimum at the origin (using \verb|SubhaloPos| from the {\sc subfind} catalogs).
We then define two characteristic radii: $R_{\rm in}$, the radius enclosing 5\% of the star formation in the galaxy, and $R_{\rm out}$, the distance enclosing 90\% of star formation within 10 kpc.
We rotate the galaxy to the face-on orientation by computing the direction of the angular momentum vector of the galaxy via all star forming cells within $R_{\rm in} < r < R_{\rm out}$ (i.e., the star-forming disk).
We then orient the system so that this vector is pointing in the $+\hat{z}$ direction.
Next, we construct 2D mass-weighted metallicity maps of each of the rotated galaxies.
We use maps with ``pixels'' $0.5~{\rm kpc}\times0.5~{\rm kpc}$.\ignorespaces
\footnote{\ignorespaces
We note that this pixel resolution was chosen to accommodate the SIMBA mass and spatial resolutions. 
Creating smaller pixels is possible in EAGLE, Illustris, and TNG; however, we confirm that varying the pixel size to smaller values (i.e., $0.25~{\rm kpc}\times0.25~{\rm kpc}$ or $0.1~{\rm kpc}\times0.1~{\rm kpc}$) does not significantly impact the core results in these models.
}
We remove pixels in these maps with gas surface densities of $\Sigma_{\rm gas} < 10^6 M_\odot\,{\rm kpc}^{-2}$ (as these regions of low density gas are unlikely to contain star-forming regions; \citeauthor{Ma_2017} \citeyear{Ma_2017}) and deproject the 2D map into a radial profile (as in the gray background histograms of the bottom panels of Figure~\ref{fig:methods_fig}).

We further reduce the radial profile into a single median relation following directly from \cite{Hemler_2021}.
We generate a median profile in bins of 0.1 kpc by searching a region of $\Delta r = \pm 0.25$ kpc for 4 populated pixels.
If the required number of valid pixels are not found, we increase $\Delta r$ to $\pm0.5$ kpc\edit{, and then $\pm1.0$ kpc}.
If this criteria is not satisfied within \edit{$\pm1$} kpc, that radial bin is removed.
The result of this median profile fitting is shown as the blue line in Figure~\ref{fig:methods_fig}.

Finally, we calculate the metallicity gradient using this median metallicity profile.
We fit a region of $R^\prime_{\rm in} < r <R_{\rm out}$ (where $R^\prime_{\rm in} = R_{\rm in} + 0.25 [R_{\rm out} - R_{\rm in}]$) which we henceforth refer to as the ``star forming region'' of the galaxy.
Not fitting the inner quarter of the star forming disk avoids the central regions of the galaxy that deviate significantly from the outer three-quarters, which can be seen in both bottom panels of Figure~\ref{fig:methods_fig} and follows directly from previous work \citep{Pilkington_2012,Gibson_2013,Ma_2017,Hemler_2021}.
Moreover, we stop the fitting at $R_{\rm out}$ as the outer regions also deviate significantly from the star forming disk.
Specifically, gradients (both in simulations and observations) seem to flatten out at large galactocentric radii \citep{Bresolin_2009,Sanchez_2014,Belfiore_2016,Sanchez_Menguiano_2018,Grasha_2022,Tapia_2022,Chen_2023, Garcia_2023}.
Thus, we fit the median profile only in the star forming region with a single linear regression (red line in Figure~\ref{fig:methods_fig}).
The slope of this line is what we use as the metallicity gradient of the system.

We note that our method of calculating the metallicity gradient within the star forming region as we have defined it is not the only way to do so.
Appendix~\ref{appendix:gradient_definitions} therefore includes a comparison of our sample over a different gradient region: $0.5R_{\star} < r < 2.0R_{\star}$ (where $R_{\star}$ is the stellar half mass radius).
Briefly, we find that galaxies' measured gradient is roughly unchanged in EAGLE and SIMBA with a varied gradient definition.
At high redshift ($z\geq4$) in Illustris and virtually all redshifts in TNG, the star forming region cut used in the main body of this work tends to exhibit flatter gradients.
Taking this fixed size at higher redshifts is less common in observational literature, however, owing to the limited resolution at these higher redshifts \citep[see, e.g.,][who make no such size cut]{Troncoso_2014,Venturi_2024}.
We therefore opt to use the gradient definition that should most closely track with the star forming gas: that is, the gas that is most likely to have emission lines from which a metallicity can be determined.
We discuss this in more detail in Appendix~\ref{appendix:gradient_definitions}.

We also require that the star forming region be at least 1 kpc (in the 1D profile).
We note that 1 kpc is potentially large for galaxies at very high redshift \citep{Ormerod_2024}; however, even by relaxing this assumption, we find that galaxies below this size tend to fail other selection criteria.
This criterion is therefore made for practical reasons.
Higher resolution simulations are needed to characterize galaxies with more compact star-forming regions.
Finally, we require the median profile to be mostly contiguous by demanding that there be valid data covering at least 90\% of the star forming region.

The redshift ranges of each simulation sample are not necessarily consistent owing to the application of the above methodology on varying mass and spatial resolutions.
In particular, we report the distribution of gradients in EAGLE at $0\leq z\leq7$, Illustris at $0\leq z\leq8$, TNG at $0\leq z\leq8$, and SIMBA at $0\leq z\leq4$.

\section{Results}
\label{sec:results}

\subsection{The Redshift Evolution of Metallicity Gradients}
\label{subsec:each_sim_gradients}

\begin{figure*}
    \centering
    \includegraphics[width=\linewidth]{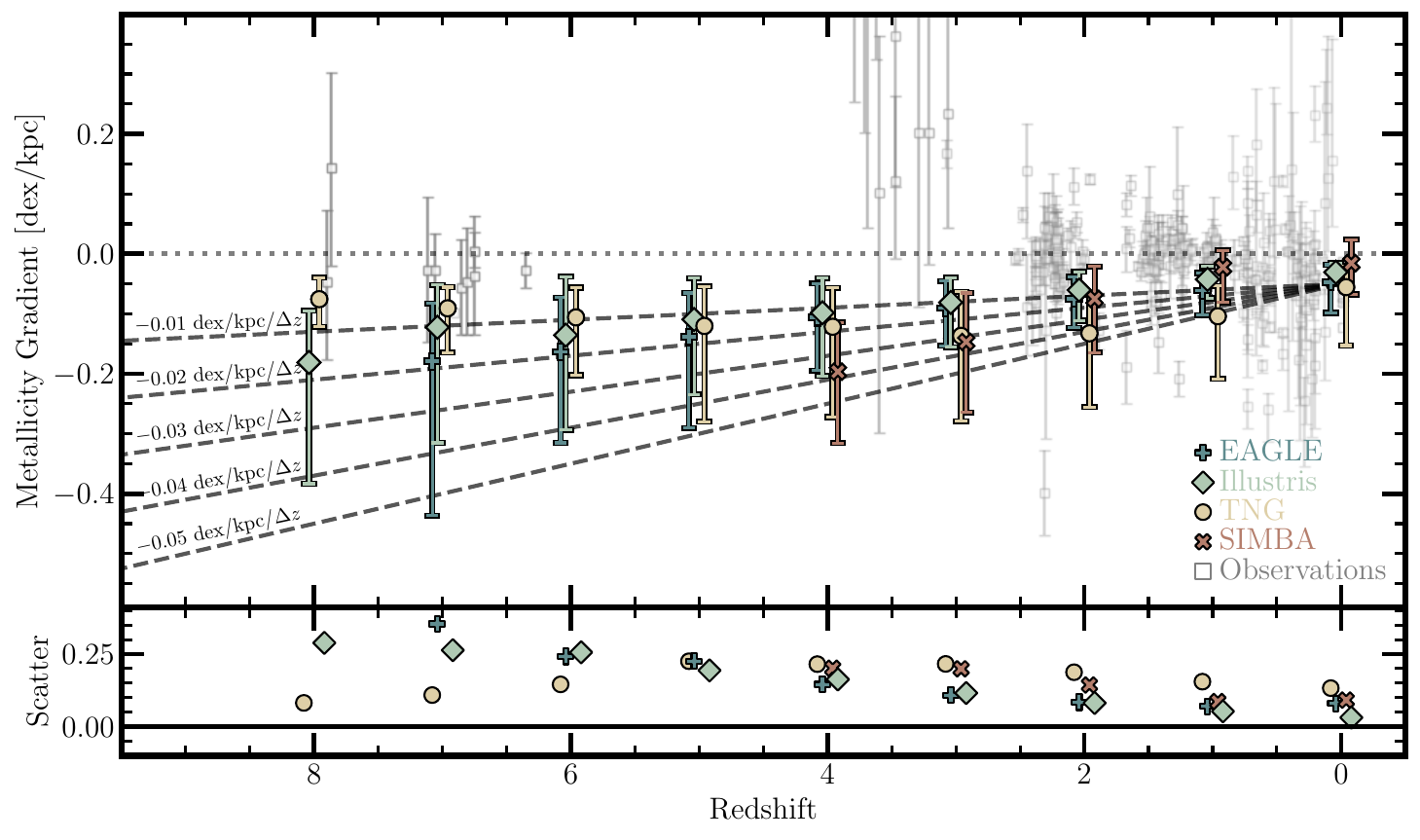}
    \caption{
    {\bf The Redshift Evolution of Metallicity Gradients in EAGLE, Illustris, TNG, and SIMBA.}
    {\it Top:} The median, $16^{\rm th}$ and $84^{\rm th}$ percentile of the gradient distributions in EAGLE (plus), Illustris (diamond), TNG (circle), and SIMBA (x).
    We also include evolutionary lines of $-0.01$, $-0.02$, $-0.03$, $-0.04$, and $-0.05$ ${\rm dex/kpc/}\Delta z$ (with intercepts of $-0.05~{\rm dex/kpc}$) as a point of reference.
    The observed metallicity gradients (unfilled squares) are a collection of \protect\cite{Rupke_2010b,Queyrel_2012,Swinbank_2012,Jones_2013,Jones_2015,Troncoso_2014,Leethochawalit_2016,Wang_2017,Wang_2019,Wang_2022,Carton_2018,Forster_Schreiber_2018,Curti_2020,Grasha_2022,Li_2022,Arribas_2024,Ju_2024,Vallini_2024,Venturi_2024}.
    We emphasize that the errorbars on the observations are uncertainty on individual gradient measurements, whereas the errorbars on the simulation points are the spread of the distributions.
    Additionally, we note that the simulation data points are slightly offset from their respective redshifts for aesthetic purposes; however, all simulation data points are at integer redshifts.
    {\it Bottom:} The scatter about the median for each redshift in each simulation, defined as magnitude of difference between $84^{\rm th}$ and $16^{\rm th}$ percentiles.
    We note that the median, $16^{\rm th}$ percentile, and $84^{\rm th}$ percentile for each simulation can also be found in Table~\ref{tab:properties_1}.
    }
    \label{fig:summary_fig}
\end{figure*}

\begin{table*}
    \centering
    \begin{tabular}{x{0.1\linewidth}rx{0.125\linewidth}rx{0.125\linewidth}rx{0.125\linewidth}rx{0.125\linewidth}}
\toprule
& \multicolumn{2}{c}{EAGLE} & \multicolumn{2}{c}{Illustris} & \multicolumn{2}{c}{TNG} & \multicolumn{2}{c}{SIMBA}\\\cmidrule(lr){2-3}\cmidrule(lr){4-5}\cmidrule(lr){6-7}\cmidrule(lr){8-9}
& \multicolumn{1}{c}{$N$} & $\nabla$ [dex/kpc] & \multicolumn{1}{c}{$N$} & $\nabla$ [dex/kpc] & \multicolumn{1}{c}{$N$} & $\nabla$ [dex/kpc] & \multicolumn{1}{c}{$N$} & $\nabla$ [dex/kpc] \\\midrule
$z=0$ & \lc{$5851$} & $-0.047_{-0.098}^{-0.018}$ & \lc{$19101$} & $-0.031_{-0.047}^{-0.015}$ & \lc{$3507$} & $-0.056_{-0.153}^{-0.020}$ & \lc{$389$} & $-0.016_{-0.068}^{+0.024}$\\
$z=1$ & \lc{$7673$} & $-0.061_{-0.102}^{-0.032}$ & \lc{$16105$} & $-0.043_{-0.074}^{-0.021}$ & \lc{$4675$} & $-0.104_{-0.209}^{-0.053}$ & \lc{$342$} & $-0.022_{-0.081}^{+0.006}$\\
$z=2$ & \lc{$5762$} & $-0.075_{-0.123}^{-0.039}$ & \lc{$10364$} & $-0.061_{-0.112}^{-0.030}$ & \lc{$4661$} & $-0.133_{-0.256}^{-0.068}$ & \lc{$167$} & $-0.075_{-0.165}^{-0.021}$\\
$z=3$ & \lc{$3087$} & $-0.091_{-0.153}^{-0.045}$ & \lc{ $5557$} & $-0.081_{-0.156}^{-0.040}$ & \lc{$3651$} & $-0.137_{-0.279}^{-0.063}$ & \lc{ $77$} & $-0.147_{-0.265}^{-0.065}$\\
$z=4$ & \lc{$1432$} & $-0.106_{-0.195}^{-0.049}$ & \lc{ $2591$} & $-0.098_{-0.203}^{-0.040}$ & \lc{$2317$} & $-0.122_{-0.272}^{-0.057}$ & \lc{ $29$} & $-0.197_{-0.316}^{-0.114}$\\
$z=5$ & \lc{ $461$} & $-0.139_{-0.291}^{-0.065}$ & \lc{  $982$} & $-0.110_{-0.234}^{-0.041}$ & \lc{$1243$} & $-0.120_{-0.280}^{-0.054}$ & \lcTwo{~}\\
$z=6$ & \lc{ $150$} & $-0.164_{-0.315}^{-0.073}$ & \lc{  $325$} & $-0.136_{-0.294}^{-0.038}$ & \lc{ $583$} & $-0.106_{-0.202}^{-0.057}$ & \lcTwo{~}\\
$z=7$ & \lc{  $34$} & $-0.179_{-0.437}^{-0.083}$ & \lc{   $82$} & $-0.123_{-0.315}^{-0.052}$ & \lc{ $230$} & $-0.091_{-0.165}^{-0.055}$ & \lcTwo{~}\\
$z=8$ & \lcTwo{}                                 & \lc{   $14$} & $-0.181_{-0.384}^{-0.095}$ & \lc{  $73$} & $-0.076_{-0.122}^{-0.040}$ & \lcTwo{}\\\hline
dex/kpc/$\Delta z$ & \lcTwo{$-0.015 \pm 0.001$} & \lcTwo{$-0.016 \pm 0.001$} & \lcTwo{$-0.016 \pm 0.004$} & \lcTwo{$-0.028 \pm 0.007$} \\
\bottomrule
    \end{tabular}
    \caption{{\bf Average Gradients in the Samples}.
    The number of galaxies ($N$) and median metallicity gradient ($\nabla$) for every redshift where there are more than 10 galaxies that pass our selection criteria (outlined in Section~\ref{subsec:selection_criteria}). 
    The subscripts and superscripts on the median gradients are the $16^{\rm th}$ and $84^{\rm th}$ percentiles of the distributions.
    The bottom row shows the average evolution of the gradients in each simulation (regression via medians weighted by the number of galaxies in each bin).
    The quoted uncertainties are the square root of the variance of the slope, taken from the covariance matrix.
    These data are also shown graphically in Figure~\ref{fig:summary_fig}.
    }
    \label{tab:properties_1}
\end{table*}

The top panel of Figure~\ref{fig:summary_fig} shows the distribution of metallicity gradients in each simulation as a function of redshift.
We characterize the distributions with their median and their spread with the $16^{\rm th}$ and $84^{\rm th}$ percentiles.
We note that we require there to be at least 10 galaxies that pass our selection criteria to report the distribution of galaxies.
These data are also presented in Table~\ref{tab:properties_1} for ease of reference, along with the number of galaxies in each redshift bin that pass the selection criteria.
\edit{We, again, caution that the distribution of galaxies presented here are {\it not} individual galaxy gradient evolution traced through time.
The distributions of galaxies presented here represents all galaxies that pass our selection criteria at that given redshift.
}
We \edit{also} note that \citet{Hemler_2021} quote the peak of a log normal distribution instead of a median.
At $z\sim0$, the distributions are indeed fairly well characterized by a log normal; however, at higher redshift the distributions become significantly less log normally distributed.
The net effect of this is that the median is at slightly steeper values than the peak of the log normal distributions.
Figure~\ref{fig:summary_fig} also has a compilation of observed metallicity gradients (gray squares), we make direct comparisons to these gradients in Section~\ref{subsec:comparison_with_obs}.

Generally speaking, each of the simulations has similar behavior: increasingly negative gradients with increasing redshift.
This evolution is roughly linear in redshift space in EAGLE, Illustris, and SIMBA.
In TNG, on the other hand, there is a plateau at $z=4-5$ before the gradients start to become flatter towards $z=8$.
We obtain gradient evolutions of $-0.015\pm0.01$ dex/kpc/$\Delta z$, $-0.016\pm0.001$ dex/kpc/$\Delta z$, $-0.016\pm0.004$ dex/kpc/$\Delta z$, and $-0.028\pm0.007$ dex/kpc/$\Delta z$ in EAGLE, Illustris, TNG, and SIMBA, respectively, by fitting a linear regression to the median gradients at each redshift (weighted by the number of galaxies in each redshift bin; see the bottom row of Table~\ref{tab:properties_1})\ignorespaces
\footnote{\ignorespaces
We note that we perform the regression on the medians weighted by the number of galaxies in each mass bin instead of through every gradient individually to avoid the least squares algorithm from over-fitting to outliers (see Appendix~\ref{appendix:regression}).
}.

The bottom panel of Figure~\ref{fig:summary_fig} shows the scatter of each distribution (taken as the difference between the $84^{\rm th}$ and $16^{\rm th}$ percentiles of the distributions).
EAGLE, Illustris, and SIMBA share the most similar trends in the scatter about their distributions.
In EAGLE, the scatter increases roughly linearly with increasing redshift.
The scatter at $z=0$ is $0.08~{\rm dex/kpc}$ and increases to $>0.3~{\rm dex/kpc}$ at $z=7$.
So, too, does the scatter in Illustris increase roughly linearly with increasing redshift starting quite small at $\sim0.03~{\rm dex/kpc}$ to $\sim0.25~{\rm dex/kpc}$ at $z=8$.
Finally, SIMBA follows the pattern with a roughly linearly increasing scatter with increasing redshift.
The $16^{\rm th}$ percentile of the distribution in particular changes more significantly than the $84^{\rm th}$ percentile, suggesting there are more steep negative gradients at higher redshift than lower in EAGLE, Illustris, and SIMBA.
TNG breaks the pattern by increasing linearly with increasing redshift to around $z\sim4$ (around $0.25~{\rm dex/kpc}$), plateauing to $z=5$, and then decreasing back to $z=8$ (to a small $0.08~{\rm dex/kpc}$).
This behavior is qualitatively very similar to that of the overall gradients in TNG, which become steeper back to $z=4$, plateau to $z=5$, and then become flatter back to $z=8$.
It is likely that this behavior, both in the scatter and the medians, is a result of the mass distribution of the sample changing with increasing redshift, as we will discuss in more detail in the next section.

It should be remarked that, for the most part, the qualitative behavior of the simulation models is really quite similar, despite the quantitative differences.
The level of qualitative agreement is, perhaps, expected based on previous results \citep[see, e.g.,][]{Gibson_2013}.
Each of the models of this work employs an ISM treatment that gives rise to smooth stellar feedback (see Section~\ref{subsec:simulations}) which allows for the persistence of the radially decreasing metallicity profiles through cosmic time.

\subsection{Gradient Evolution By Mass}
\label{subsection:evolution_by_mass}

\begin{figure*}
    \centering
    \includegraphics[width=\linewidth]{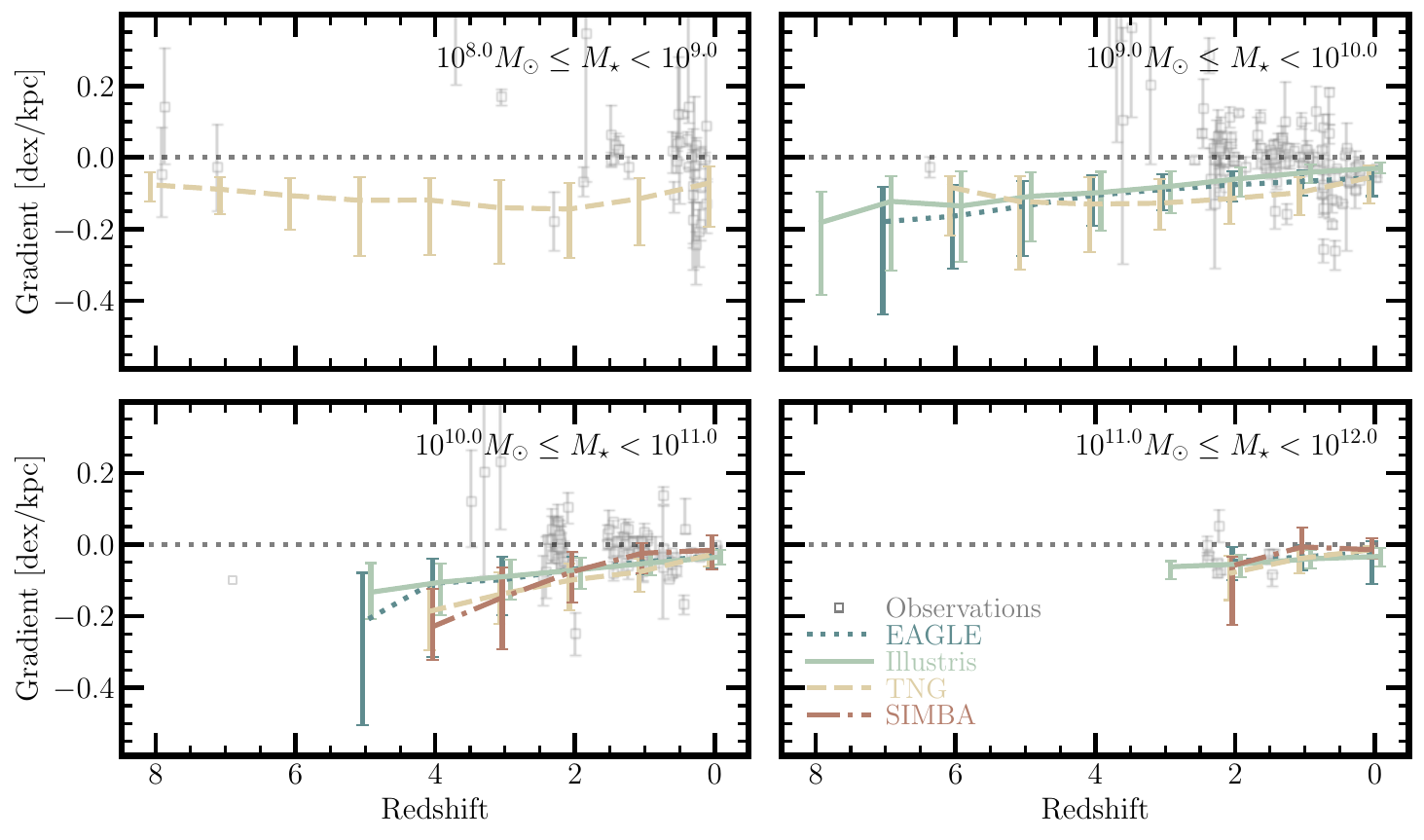}
    \caption{{\bf Metallicity Gradient Evolution by Stellar Mass.}
    The gradient evolution of star forming galaxies in a combination of EAGLE (dotted), Illustris (solid), TNG (dashed), and SIMBA (dot dash) broken into four stellar mass bins of width 1.0 dex ranging from $10^{8.0}~M_\odot$ to $10^{12.0}~M_\odot$ as labeled in each panel.
    The errorbars represent the width of the distributions ($16^{\rm th}$ and $84^{\rm th}$ percentiles).
    The gray data points are \edit{from the same set of observations as in Figure~\ref{fig:summary_fig} (when masses are available in their respective references).}
    }
    \label{fig:evolution}
\end{figure*}

The direct comparison of gradients in different simulations above makes two implicit assumptions: (i) that the evolution of gradients is the same in galaxies across subpopulations and (ii) that the samples in each simulation contain the same subpopulations.
These are not necessarily the case, however.
As noted in Section~\ref{subsec:selection_criteria}, each simulation sample has a different minimum stellar mass owing to the varied mass resolutions.
To that end, this section first addresses how the gradients of galaxies in different mass bins evolve and then investigates the implications this has on the interpretation of the results of Section~\ref{subsec:each_sim_gradients}.

Figure~\ref{fig:evolution} shows the evolution of metallicity gradients of all four simulations broken down into four different stellar mass bins: $10^{8.0}M_\odot\leq M_\star < 10^{9.0}M_\odot$ (top left), $10^{9.0}M_\odot\leq M_\star < 10^{10.0}M_\odot$ (top right),
$10^{10.0}M_\odot \leq M_\star < 10^{11.0}M_\odot$ (bottom left), and $10^{11.0}M_\odot\leq M_\star < 10^{12.0}M_\odot$ (bottom right).
\edit{Each mass bin (both here and in Appendix~\ref{appendix:mass_evo_fixed_z}) represents the stellar mass of the galaxy {\it at each redshift} (i.e., not the mass of its $z=0$ descendant).
}
We first note that each simulation has qualitatively similar behavior.
Generally, we find that galaxies in the lowest mass bins ($M_\star < 10^9 M_\odot$) tend to have virtually no redshift evolution ($\sim0.00~{\rm dex/kpc/}\Delta z$).
Galaxies in the highest mass bins ($10^{11.0}M_\odot\leq M_\star < 10^{12.0}M_\odot$; bottom right) also show very little redshift evolution.
The caveat here, of course, is that there are virtually no galaxies in any of the simulations here at stellar masses of greater than $10^{11.0}M_\odot$ at $z>3$.
The extent to which this trend holds at $z>3$ is therefore not clear.
The intermediate mass bins ($10^{9.0}M_\odot\leq M_\star < 10^{10.0}M_\odot$; top right, and $10^{10.0}M_\odot \leq M_\star < 10^{11.0}M_\odot$; bottom left) have much stronger evolution.
While there are quantitative differences (which we will discuss more below), the generality of this result across the different simulation models is quite remarkable.
\edit{The generality of this result is similarly found when looking at the gradient evolution as a function of mass in fixed redshift bins (see Appendix~\ref{appendix:mass_evo_fixed_z} Figure~\ref{fig:appendix_D1}).}

Looking more quantitatively, we can fit the evolution of the gradients with a linear regression in each simulation and in each mass bin in the same way as in Section~\ref{subsec:each_sim_gradients} (shown in  Table~\ref{tab:evolution_by_mass}).
The quantitative trends confirm those of the qualitative trends: the $10^{8.0}M_\odot \leq M_\star < 10^{9.0}M_\odot$ bin has the weakest redshift evolution\footnote{\ignorespaces
It should be noted though that there is a slight `dip' in the evolution of the gradients at $z\sim2-3$ in the $10^{8.0}M_\odot \leq M_\star < 10^{9.0}M_\odot$ bin.
This implies that galaxies of this mass should have their steepest negative gradients at $z\sim2-3$.
\label{footnote:dip}
},
$10^{10.0}M_\odot \leq M_\star < 10^{11.0}M_\odot$ bin has the strongest, with the other two mass ranges having moderate evolution.
Moreover, this trend is qualitatively similar as that of \cite{Belfiore_2017} at $z\sim0$ in SDSS, who find that gradients are the most negative in galaxies with stellar masses of $10^{10.0}M_\odot<M_\star<10^{10.5}M_\odot$ (see, e.g., their Figure B1).
Although, it should be noted that in the simulations at $z=0$ galaxies with stellar masses ranging from $10^{9.0}M_\odot<M_\star<10^{10.0}M_\odot$ tend to have the most negative gradients.

\begin{table*}
    \centering
    \begin{tabular}{llx{0.15\linewidth}x{0.15\linewidth}x{0.15\linewidth}x{0.15\linewidth}}
        \toprule
        & & \multicolumn{4}{c}{Stellar Mass Bin $[\log M_\odot]$} \\\cmidrule(lr){3-6}
        Simulation & & $8.0-9.0$ & $9.0-10.0$ & $10.0-11.0$ & $11.0-12.0$ \\\midrule
        EAGLE     & $[{\rm dex/kpc}/\Delta z]$ & -- & $-0.011\pm0.001$ & $-0.019\pm0.001$ & $-0.013\pm0.000$ \\
        & $[{\rm dex/}R_{\rm SFR}/\Delta z]$ & -- & $-0.006\pm0.002$ & $-0.009\pm0.002$ & $-0.001\pm0.001$ \\\midrule
        Illustris & $[{\rm dex/kpc}/\Delta z]$ & -- & $-0.016\pm0.001$ & $-0.017\pm0.000$ & $-0.009\pm0.002$ \\
        & $[{\rm dex/}R_{\rm SFR}/\Delta z]$ & -- & $-0.006\pm0.001$ & $-0.005\pm0.001$ & $-0.001\pm0.001$ \\\midrule
        TNG       & $[{\rm dex/kpc}/\Delta z]$ & $-0.007\pm0.005$ & $-0.024\pm0.004$ & $-0.037\pm0.004$ & $-0.027\pm0.005$ \\
        & $[{\rm dex/}R_{\rm SFR}/\Delta z]$ & $-0.005\pm0.001$ & $-0.014\pm0.000$ & $-0.022\pm0.005$ & $-0.006\pm0.002$ \\\midrule
        SIMBA     & $[{\rm dex/kpc}/\Delta z]$ & -- & -- & $-0.030\pm0.011$ & $-0.022\pm0.021$ \\
        & $[{\rm dex/}R_{\rm SFR}/\Delta z]$ & -- & -- & $-0.018\pm0.007$ & $-0.015\pm0.012$ \\
        \bottomrule
    \end{tabular}
    \caption{{\bf Average Redshift Evolution of Metallicity Gradients By Mass.}
    The best-fit linear regression parameters (weighted by number of galaxies) to the evolution of gradients for every simulation analyzed in this work.
    The quoted uncertainties are the square root of the variance of the slope, taken from the covariance matrix.
    We find that galaxies with stellar masses of $10^{10}M_\odot \leq M_\star < 10^{11}$ tend to have the strongest ${\rm dex/kpc/}\Delta z$ gradient evolution.
    Moreover, we find that much of the redshift evolution goes away when metallicity gradient is normalized by galaxy size (${\rm dex/}R_{\rm SFR}/\Delta z$; where $R_{\rm SFR}$ is the radius enclosing 50\% of the star forming gas).
    }
    \label{tab:evolution_by_mass}
\end{table*}

\section{Discussion}
\label{sec:discussion}

\subsection{On the Mass Dependence of Metallicity Gradients in Smooth Stellar Feedback Models}
\label{subsec:mass_dependence_discussion}

\begin{figure*}
    \centering
    \includegraphics[width=\linewidth]{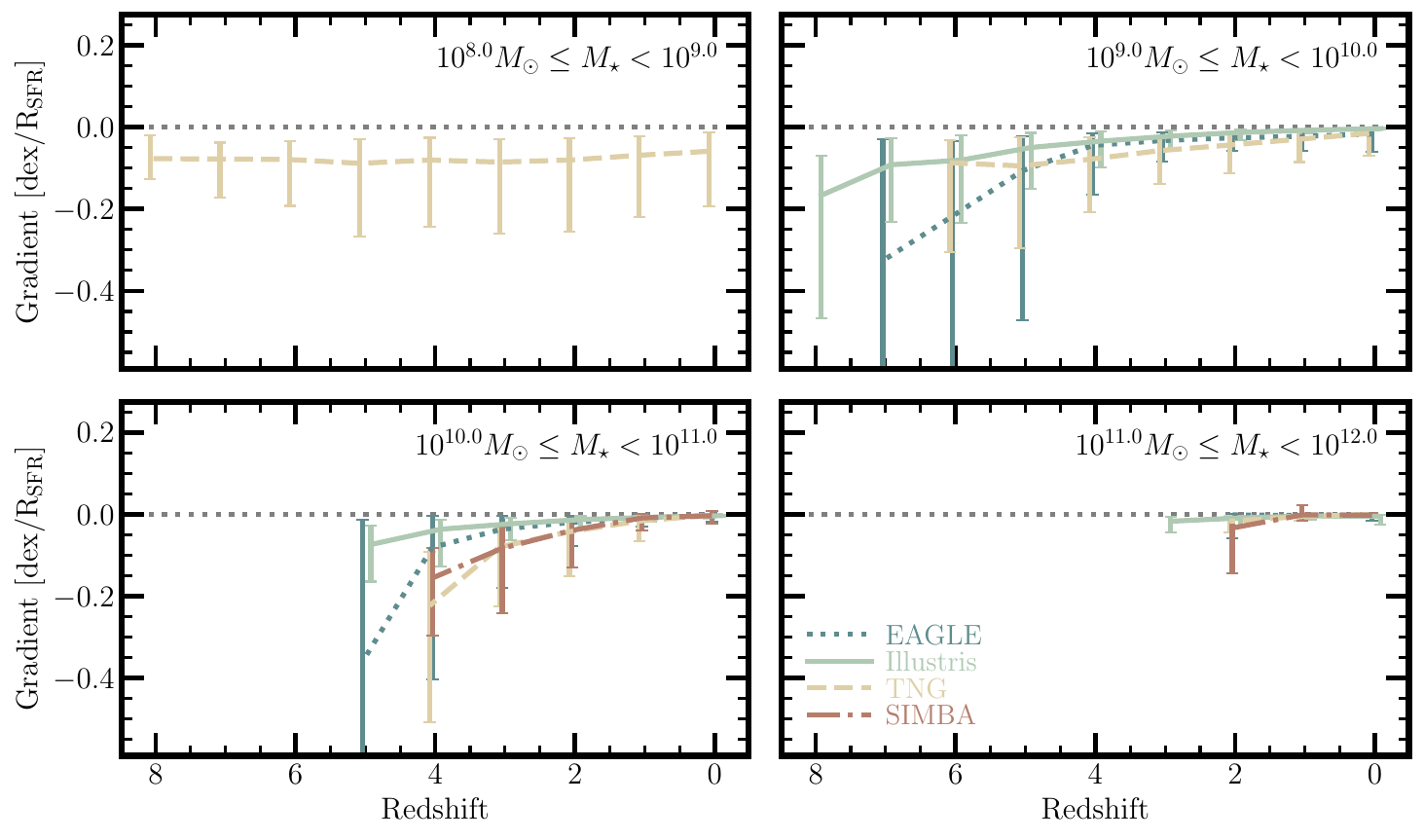}
    \caption{{\bf Metallicity Gradient Evolution by Stellar Mass Normalized by Galaxy Size.}
    Same as Figure~\ref{fig:evolution}, but with each galaxy's gradient normalized by $R_{\rm SFR}$ (the radius enclosing 50\% of the total star formation) as a proxy for the size of the galaxy.
    \edit{We note that we do not compare to observations in this figure as the normalization by size is done in a non-homogenous way in the literature.
    Direct comparisons of the normalized gradients are therefore difficult to make.
    }}
    \label{fig:normalized_by_size}
\end{figure*}

In the previous sections, we found that galaxies in simulations that employ smooth stellar feedback have similar metallicity gradient evolution, particularly when broken into stellar mass bins.
In this section, we examine what might be causing this coherent metallicity gradient-stellar mass evolution.

Figure~\ref{fig:normalized_by_size} shows the metallicity gradients from EAGLE, Illustris, TNG, and SIMBA (broken down into the same mass bins as Figure~\ref{fig:evolution}) normalized by a proxy for the size of the galaxy.
The size we choose is $R_{\rm SFR}$, the radius enclosing half of the total star formation within a galaxy \citep[as in][]{Hemler_2021}.\ignorespaces
\footnote{\ignorespaces
We note that other reasonable choices for the size of galaxies, e.g., the stellar half-mass radius, yield similar results as seen in Figure~\ref{fig:normalized_by_size}.
}
We generally find that much of the redshift evolution in the metallicity gradients is removed in each of the simulations (evolution reduced by a factor of $\gtrsim2$; see Table~\ref{tab:evolution_by_mass}).
In the lowest mass bin ($10^{8.0}M_\odot \leq M_\star \leq 10^{9.0}M_\odot$), we find that the aforementioned ``dip'' in the strength of gradients at $z\sim2-3$ (see footnote~\ref{footnote:dip}) is reduced significantly when normalized by galaxy size.
Similarly, galaxies in the highest mass bin ($10^{11.0}\leq M_\star \leq 10^{12.0}$) have very flat metallicity gradients when normalized by size that remain very flat with time.
Galaxies with stellar masses of $10^{9.0}M_\odot \leq M_\star \leq 10^{10.0}$ in EAGLE, Illustris, and TNG have virtually no redshift evolution in their gradients from $z=0-4$ when normalized by galaxy size.
Past $z\sim4$, however, the metallicity gradients seem to become more negative.
A similar behavior is seen in the $10^{10.0}M_\odot\leq M_\star\leq 10^{11.0}M_\odot$ bin: gradients are very flat with no evolution at $z\lesssim2$, but then start to become more negative with increasing redshift, albeit at different rates in the simulation models.
Regardless of the exact details, it is once again worth noting how remarkably well these simulation models agree.

We suggest that the evolution of metallicity gradients in these models is therefore mostly governed by the size of the galaxies.
The smooth feedback models allow galaxies to assemble their metal content in the canonical inside-out fashion.
In the absence of any large outflows and/or small-scale metal mixing -- neither of which should be expected from models of this type (see also discussion in Section~\ref{subsec:comparison_with_obs}) -- the initial build-up of metals should persist in the galaxy through time.
As time progresses, the physical extent of galaxy increases leading to an increased ${\rm dex/kpc}$ gradient but maintaining its ${\rm dex/size}$ gradient.
Indeed, the qualitative picture of a coherent evolution in galaxy size and metallicity gradient has been noted previously in both simulations \citep{Tissera_2019,Hemler_2021} and in observations \citep{Sanchez_2012,Sanchez_2014,Ho_2015,Sanchez_Menguiano_2016,Sanchez_Menguiano_2018}.

Of course, the increasingly negative gradients at high redshift in the $10^{9.0}M_\odot\leq M_\odot\leq 10^{10.0}M_\odot$ and $10^{10.0}M_\odot\leq M_\odot\leq 10^{11.0}M_\odot$ bins complicate the picture.
The most likely source of the high redshift deviations is that these models (with the exception of Illustris) all have some implicit redshift evolution in their stellar feedback driven winds (see Section~\ref{subsec:simulations} and references therein for the full discussion).
More concretely:
(i) the EAGLE wind model depends on metallicity that allows for more efficient feedback at high redshift,
(ii) the TNG wind model has a minimum wind speed meant to produce more efficient feedback at high redshift, and 
(iii) SIMBA manually decreases mass loading factors at $z>3$ to allow for more efficient galaxy growth.
It is, perhaps, not surprising that there is some redshift dependence to the evolution of metallicity gradients in ${\rm dex/}R_{\rm SFR}$, then.
It is also suggestive that the strong redshift evolution of ${\rm dex/}R_{\rm SFR}$ gradients is weakest in Illustris, which has no explicit redshift scaling to the stellar feedback.

Finally, it should be noted that the exact normalization of the gradients in ${\rm dex/}R_{\rm SFR}$ is not the same between the four mass bins, notwithstanding the high redshift deviations.
Generally, galaxies with higher masses tend to have slightly flatter ${\rm dex/}R_{\rm SFR}$ gradients (though this effect is subtle).
One possible explanation for this is the increase in active galactic nuclei (AGN) feedback for higher mass galaxies in the simulations (see, \citeauthor{Booth_2009} \citeyear{Booth_2009} for EAGLE; \citeauthor{Vogelsberger_2013} \citeyear{Vogelsberger_2013} for Illustris; \citeauthor{Weinberger_2017} \citeyear{Weinberger_2017} for TNG; \citeauthor{Angles_Alcazar_2017a} \citeyear{Angles_Alcazar_2017a} for SIMBA).
While a detailed examination on the effects of AGN feedback on the galaxies is out of the scope of this work, it is possible that increased AGN feedback could be acting as another mixing agent in the ISM flattening metallicity gradients in higher mass galaxies.

\subsection{On The Apparent Tension in the Literature Between TNG and EAGLE}

As mentioned in Section~\ref{sec:intro}, there is an apparent tension in the literature between TNG \citep{Hemler_2021} and EAGLE \citep{Tissera_2022}.
To briefly reiterate the point of contention, \cite{Hemler_2021} find that gradients in TNG become more negative further back in time (at a rate of $\sim-0.02~{\rm dex/kpc/}\Delta z$) while \cite{Tissera_2022} find that gradients in EAGLE have very little redshift evolution\ignorespaces
\footnote{\ignorespaces
We note that \cite{Tissera_2019} use the $0.5R_\star < r < 2.0R_\star$ metallicity gradient method, whereas \cite{Hemler_2021} use the star forming region gradient.
The tension between the two papers is very likely {\it not} caused by this difference in methodology, however.
We show in Appendix~\ref{appendix:gradient_definitions} that the $0.5R_\star < r < 2.0R_\star$ gradient is virtually identical to that of the star forming region gradient in EAGLE (see left-most panel of Figure~\ref{fig:appendix_method_comparison}).
}.
Interestingly, we do not find a similar tension in this work.
The \cite{Hemler_2021}-\cite{Tissera_2022} tension is unexpected as the implementation of the star forming ISM in the EAGLE and TNG models (see Section~\ref{subsec:simulations}) which give rise to smooth feedback histories.
The qualitatively similar feedback\ignorespaces
\footnote{\ignorespaces
We note that while the implementations of feedback in EAGLE and TNG are fairly different (see Sections~\ref{subsubsec:eagle}~and~\ref{subsubsec:TNG}, respectively), the star formation histories -- and treatment of the ISM with an equation of state -- are more similar to those of high resolution simulation with strong feedback bursts (e.g., FIRE, \citeauthor{Muratov_2015} \citeyear{Muratov_2015}; \citeyear{Muratov_2017}).
}
in EAGLE and TNG should nominally lead to similar metallicity gradients \citep[and evolution thereof, see, e.g.,][]{Gibson_2013}.
We argue that the tension is due to the different population of galaxies within the samples, rather than the detailed methodological choices.

\cite{Tissera_2019} use a $(25~{\rm cMpc})^3$ volume of EAGLE (the {\sc recalL025N0725} model).
\cite{Hemler_2021}, on the other hand, use the larger TNG50 box.
Naturally, this means that the sample of galaxies in the smaller EAGLE box will be shifted towards lower masses (in the case of \citeauthor{Tissera_2022} \citeyear{Tissera_2022}, towards $M_\star\sim10^{9.0}M_\odot$).
However, lower-mass bins tend to have a more modest redshift evolution in these simulations (see Figure~\ref{fig:evolution}).
Thus, by preferentially omitting higher mass galaxies, the evolution of the gradients should be expected to be slightly flatter in the small EAGLE box.
The effect of sample selection can also be demonstrated by examining the different size TNG boxes.
Appendix~\ref{appendix:resolution} shows the metallicity gradient evolution in 5 different TNG resolution and box-size variations (TNG50-2, TNG50-3, TNG100-1, TNG100-2, and TNG300-3).
We find that, regardless of the resolution or box size of the simulation, the dominant factor driving the observed ${\rm dex/kpc}$ metallicity gradient evolution in the models is the sample of galaxies.

It should be noted that there are other subtle distinctions that give rise to some differences.
Indeed, we find that there is some level of inherent variation between the results of EAGLE and TNG (as well as Illustris and SIMBA).
The reason for this likely comes down to the different implementations of star formation, chemical enrichment, and/or stellar feedback between EAGLE and TNG (see Sections~\ref{subsubsec:eagle}~and~\ref{subsubsec:TNG}, respectively, and references therein).

Regardless, this apparent literature tension highlights the need for robust comparisons between multiple simulation models\ignorespaces
\footnote{\ignorespaces
In addition, analysis with multiple realizations of the model would be even more ideal; see, e.g., recent efforts by the CAMELS (\citeauthor{CAMLES_2021} \citeyear{CAMLES_2021}), DREAMS (\citeauthor{DREAMS_2024} \citeyear{DREAMS_2024}), and SPICE \citep{Bhagwat_2024} projects.
}
in single works.
This practice has become increasingly more common recently \citep[see][etc]{Garcia_2024a,Garcia_2024b,Garcia_2025,Lagos_2024,Munn_2024,Wright_2024,Puerto_Sanchez_2025} and can be used to help explain any discrepancies between nominally similar models.

\subsection{Comparison with Observations}
\label{subsec:comparison_with_obs}

As a point of comparison, we also include the observational measurements of metallicity gradients in Figure~\ref{fig:summary_fig} \citep[unfilled gray squares; data from][]{Rupke_2010b,Swinbank_2012,Jones_2013,Jones_2015,Troncoso_2014,Leethochawalit_2016,Wang_2017,Wang_2019,Wang_2022,Carton_2018,Curti_2020,Grasha_2022,Arribas_2024,Ju_2024,Vallini_2024,Venturi_2024}.
Broadly speaking, the observed metallicity gradients fall into three different categories:
(i) a wide diversity of gradients at $z\lesssim2.5$ \citep{Rupke_2010b,Swinbank_2012,Jones_2013,Jones_2015,Leethochawalit_2016,Wang_2017,Wang_2019,Carton_2018,Curti_2020,Grasha_2022,Ju_2024},
(ii) positive gradients at $z\approx3-4$ \citep{Troncoso_2014,Wang_2022} and
(iii) mostly negative, albeit relatively flat, gradients at $z>6$ \citep{Arribas_2024,Vallini_2024,Venturi_2024}.
Notably, there are presently no measured metallicity gradients in the range $4\lesssim z\lesssim6$.
In the following subsections, we make detailed comparisons between the simulations and observations at $0<z<4$ and then at $z>6$.

We first caution that the metallicity values measured in these simulations are mass-weighted, whereas those of observations are light-weighted.
To help accommodate for this difference, we required all valid pixels in our 2D maps construction to have significant gas content that is likely to host star forming gas (see Section~\ref{subsec:gradient_definitions} and \citeauthor{Ma_2017} \citeyear{Ma_2017}).
The pixels we obtain are therefore all would likely contain the emission lines required to obtain metallicities; however, they are not weighted by this factor in our analysis.

\subsubsection{Comparison with Observed Metallicity Gradients at \texorpdfstring{$0<z<4$}{0<z<4}}
\label{subsubsec:3<z<4}

In the local Universe ($z\sim0$), the observed metallicity gradients are typically negative with values of $\sim-0.05~{\rm dex/kpc}$ \citep{Rupke_2010b,Belfiore_2017,Grasha_2022}.
The results from the various simulations presented here are well in line with the observations.
Indeed, the median gradients in EAGLE and TNG at $z=0$ ($-0.047~{\rm dex/kpc}$ and $-0.056~{\rm dex/kpc}$, respectively) are remarkably good matches.
However, it has also been observed that the strength of the negative gradients does seem to depend on the mass of the galaxies \citep{Belfiore_2017}.
In particular, galaxies with stellar masses of $10^{10.0}M_\odot<M_\star<10^{10.5}M_\odot$ have the strongest negative gradients.
The mass dependence is seen in the simulations, too, albeit with a slight shift where the most negative gradients tend to be in slightly lower mass bins (i.e., $10^{9.0}M_\odot<M_\star<10^{10.0}M_\odot$).

At increasing redshift, observed galaxies tend to have a wider diversity of negative, flat, and positive metallicity gradients \citep{Swinbank_2012,Jones_2013,Jones_2015,Leethochawalit_2016,Wang_2017,Wang_2019,Forster_Schreiber_2018,Curti_2020,Li_2022, Dutta_2024, Ju_2024}.
In the range $0<z\lesssim1$ gradients are still predominantly negative \citep{Swinbank_2012,Carton_2018}.
The metallicity gradients from the four simulation models are broadly similar to the observations in this redshift range, although it should be noted that the simulations do not produce as many positive gradients.
At even higher redshifts ($1\lesssim z\lesssim2.5$), the observational trend persists with even more gradients being either flat or positive \citep[][although strong negative gradients have been observed, e.g., \citeauthor{Jones_2013} \citeyear{Jones_2013}; \citeauthor{Wang_2017} \citeyear{Wang_2017}; \citeauthor{Forster_Schreiber_2018} \citeyear{Forster_Schreiber_2018}]{Queyrel_2012,Swinbank_2012,Jones_2015,Leethochawalit_2016,Wang_2017,Curti_2020,Li_2022,Ju_2024}.
In contrast to this, the simulated gradients ubiquitously become more and more negative from $z=1-3$.
This tension was highlighted in \cite{Hemler_2021} between TNG and observations, but it appears to be a more general prediction of the smooth stellar feedback models.

Interestingly, all of the galaxies measured in the literature thus far at $z=3-4$ show positive gradients across a wide range of stellar masses \citep[$10^{8.6}M_\odot \lesssim M_\star \lesssim 10^{11.0}M_\odot$,][]{Cresci_2010,Troncoso_2014,Wang_2022}.
These positive gradients stand in stark contrast to the models in this work.
We caution that we are not claiming there are {\it no} positive gradients in these simulations at these redshifts.
In fact, at $z=3$, $77$ ($2.41$\%) gradients are positive in EAGLE, $31$ ($0.56$\%) in Illustris, $18$ ($0.49\%$) in TNG, and $2$ ($1.60$\%) in SIMBA.
At $z=4$, $47$ ($3.01$\%) gradients are positive in EAGLE, $23$ ($0.88$\%) in Illustris, $10$ ($0.43\%$) in TNG, and $2$ ($3.64$\%) in SIMBA.
There is, however, a clear qualitative difference in the sample of observed metallicity gradients at $z=3-4$ from metallicity gradients of these simulations.

\edit{
Figure~\ref{fig:evolution} also compares these same observed gradients (when mass estimates are available) to the simulated samples.
The discrepancy between simulated metallicity gradients and those of observations is especially clearly seen when broken into stellar mass bins.
Observed galaxies in the lowest two mass bins ($10^{8.0}M_\odot \leq M_\star < 10^{10.0}M_\odot$) tend to have shallow negative gradients at $z=0$ which transition to highly positive gradients up to $z\sim4$.
This increasingly positive transition in observations is the exact opposite trend seen in the simulated galaxy samples.
The tension is perhaps less significant, yet still present at higher masses ($10^{10.0}\leq M_\star <10^{12.0}$).
The observed sample of gradients appears to consistently range from shallow negative gradients to slightly positive gradients at $z\lesssim3$ for both the highest mass bins.
This is broadly consistent with the highest mass bin ($10^{11.0}M_\odot\leq M_\star<10^{12.0}M_\odot$) of the simulations.
However, the $10^{10.0}M_\odot \leq M_\star< 10^{11.0}M_\odot$ bin in the simulations displays increasingly negative gradients with increasing redshift.
Moreover, the $4>z>3$ gradients from \cite{Troncoso_2014} at $10^{10.0}M_\odot\leq M_\star<10^{11.0}M_\odot$ are significantly more positive than the simulations.
}

Although the tension at $1<z<4$ seems quite significant, it should be noted that there are some observational systematics to be taken into account in these gradients.
Several studies have examined the impact of systematically degrading the quality of high spatial resolution, high signal-to-noise data \citep{Yuan_2013,Mast_2014,Poetrodjojo_2019,Acharyya_2020,Grasha_2022,Metha_2024}.
Outside of lensed systems, there are typically only a few resolution elements per galaxy, even with the best telescopes.
Generally, metallicity gradients are found to systematically flatten with lowered angular resolution and cannot be meaningfully constrained with sufficiently low signal-to-noise ratio.
Lower spatial resolution observations can ``smear'' measurements of a single \textsc{Hii} region with either nearby \textsc{Hii} regions or diffuse ionized gas (DIG; \edit{\citeauthor{Poetrodjojo_2019} \citeyear{Poetrodjojo_2019}}).
Coarser spatial resolution that combines multiple \textsc{Hii} regions will over-weight the spectra towards the regions of stronger emission, causing an overestimation of the metallicity in the outskirts of galaxies \citep{Yuan_2013}.
DIG-dominated regions tend to have different physical conditions and emission line ratios than \textsc{Hii} regions \citep{Blanc_2009,Zhang_2017}.
Any contamination of DIG in a spectrum systematically alters the derived metallicity \citep{Poetrodjojo_2019}.
Fortunately, ongoing and future studies in the new generation of optical and infrared telescopes (i.e., JWST and future ELTs) should help alleviate some of these issues with increased spatial resolution at these high redshifts.
However, the gradients from \cite{Wang_2022} and \cite{Ju_2024} {\it do} come from JWST and are virtually all flat or positive.
The extent to which the large population of gradients can be simply explained away by poor angular resolution is therefore unclear.

Beyond the impact of observational systematics, each of the simulations employs a subgrid equation of state for the dense, star forming ISM (see Section~\ref{subsec:simulations} and references therein for more details).
An equation of state is not the unique method by which the dense ISM is modeled, however.
Simulations such as FIRE (\citeauthor{Hopkins_2014} \citeyear{Hopkins_2014}, \citeyear{Hopkins_2018}, \citeyear{Hopkins_2023}) and SMUGGLE (\citeauthor{Marinacci_2019} \citeyear{Marinacci_2019}) are higher resolution and can directly model giant molecular cloud scales ($M_{\rm baryon}\lesssim10^4 M_\odot$) of the ISM.
These treatments, known collectively as ``explicit'' ISM models, have feedback regulated star formation, which can lead to episodic blow outs of gas in a short period of time \citep[see, e.g.,][]{Muratov_2015, Muratov_2017, Angles_Alcazar_2017b, Pandya_2021}.
These episodic bursts of feedback work to systematically flatten (or even temporarily invert) metallicity gradients by rapidly redistributing material from the inner regions to the outskirts \citep{Ma_2017,Muratov_2017,Bellardini_2021,Sun_2024}.
\edit{Moreover, strong bursty feedback would preferentially impact lower mass galaxies, expelling their metals into the CGM \citep{Muratov_2017}.
The lack of strong outflows driven by strong stellar feedback could therefore account for the tension between the lower mass galaxies seen in Figure~\ref{fig:evolution}.
}
It is therefore possible that the dearth of positive gradients in this redshift range in EAGLE, Illustris, TNG, and SIMBA is indicative that subgrid ISM models do not sufficiently model galaxy feedback at $z\sim1-4$.

\subsubsection{Comparison with Observed Metallicity Gradients at \texorpdfstring{$z>6$}{z>6}}
\label{subsubsec:z>6}

Most of the gradients measured at $z>6$ are at $z\sim7$ with 7 galaxies to date (from \cite{Vallini_2024}: COS-2987 at $z=6.8$, COS-3018 at $z=6.8$, UVISTA-Z-001 at $z=7.0$, UVISTA-Z-007 at $z=6.7$, UVISTA-Z-019 at $z=6.7$; from \cite{Venturi_2024}: BDF-3299 at $z=7.1$; from \cite{Arribas_2024}: SPT0311-58 E at $z=6.9$).
Meanwhile, only one gradient has been measured at $z\sim6$ (COSMOS24108 at $z=6.3$ from \citeauthor{Venturi_2024} \citeyear{Venturi_2024}) and two at $z\sim8$ (A2744-YD4 at $z=7.8$ and A2744-YD1 at $z=7.8$ both from \citeauthor{Venturi_2024} \citeyear{Venturi_2024}).
All of the high redshift galaxies display relatively shallow negative gradients (with the lone exception of A2744-YD4 at $z=7.8$ from \citeauthor{Venturi_2024} \citeyear{Venturi_2024}, which has a strong positive gradient).
Broadly speaking, this is consistent with simulations in the redshift range, although in detail they appear slightly flatter than simulations.

We will make a more direct comparison to observed gradients at $z\sim7$ as that is the most populated redshift in the recent observations.
The median observed gradient at $z=7$ is approximately $-0.04~{\rm dex/kpc}$.
This is significantly flatter than the medians of $-0.179~{\rm dex/kpc}$, $-0.123~{\rm dex/kpc}$, and $-0.091~{\rm dex/kpc}$ in EAGLE, Illustris, and TNG at $z=7$ (respectively).
Na\"ively taking the full distributions from the simulations, the observational median of $\sim-0.03~{\rm dex/kpc}$ is outside the $84^{\rm th}$ percentiles of all of the EAGLE, Illustris, and TNG distributions.
The tension is not as strong as at $3<z<4$, but still persists.
Recall, though, that the evolution of gradients is not constant with time (see Section~\ref{subsection:evolution_by_mass}).
Lower mass galaxies tend to have flatter metallicity gradients than high mass galaxies at $z=7$, in particular.
Figure~\ref{fig:evolution} also includes these $z>6$ gradients in their respective mass bins.
Broadly speaking, these galaxies span a wide range of masses:
SPT0311-58 E (from \citeauthor{Arribas_2024} (\citeyear{Arribas_2024}) has a stellar mass of $\sim10^{10.5}M_\odot$, BDF-3299 (from \citeauthor{Venturi_2024} \citeyear{Venturi_2024}) has a stellar mass of $10^{8.21}M_\odot$, and galaxies from \cite{Vallini_2024} have masses ranging from $10^{8.0}M_\odot<M_\star<5\times10^{10}M_\odot$.
\ignorespaces
\footnote{\ignorespaces
The stellar masses of the galaxies in \cite{Vallini_2024} are not reported exactly for each galaxy.
We therefore do not report their gradients in any panel.
}
Considering the stellar masses of these systems, there is perhaps hints of a tension between the observed gradients and those of the simulations analyzed in this work, in particular with the \cite{Arribas_2024} SPT0311-58 E galaxy.
It should be noted that spatially resolved maps of galaxies at these high redshifts are highly expensive observationally.
The earliest observed targets (such as the ones we compare to here) are likely the brightest objects.
It is possible that they are atypical in some regard, perhaps, a recent merger or large gas accretion event (both of which could systematically flatten the gradient, see, e.g., \citeauthor{Rupke_2010b} \citeyear{Rupke_2010b}; \citeauthor{Torrey_2012} \citeyear{Torrey_2012}; \citeauthor{Ceverino_2016} \citeyear{Ceverino_2016}).
Larger samples of galaxies at these extreme redshifts are therefore critical for getting a statistical understanding of the extent to which the tension between observations and smooth feedback models is significant.

An additional consideration is the metallicity diagnostics at these high redshift.
\cite{Vallini_2024} use Atacama Large Millimeter/submillimeter Array (ALMA) observations using (rest) infrared lines ($[\textsc{Oiii}]$ $88\mu$m and $[\textsc{Cii}]$ $158\mu$m).
This is in contrast to the observations from JWST that use (rest) optical lines: the \cite{Curti_2017,Curti_2020b} calibrations in the case of \cite{Arribas_2024} and the \cite{Laseter_2024} $\hat{R}$ in \cite{Venturi_2024}.
There may therefore be some systematics not taken into account with the use of the rest infrared lines over the optical lines, as \cite{Vallini_2024} point out.
Moreover, it is yet to be seen as to whether the low-redshift calibrated optical line relations are valid at higher redshifts (though work has been done to calibrate these relations at high redshift, \citeauthor{Garg_2023} \citeyear{Garg_2023}; \citeauthor{Curti_2023a} \citeyear{Curti_2023a}; 
\citeauthor{Hirschmann_2023} \citeyear{Hirschmann_2023}; \citeauthor{Sanders_2023} \citeyear{Sanders_2023}; \citeauthor{Trump_2023} \citeyear{Trump_2023}; \citeauthor{Ubler_2023} \citeyear{Ubler_2023}; \citeauthor{Chakraborty_2024} \citeyear{Chakraborty_2024}, 
\citeauthor{Laseter_2024} \citeyear{Laseter_2024}, \citeauthor{Backhaus_2025} \citeyear{Backhaus_2025}).

Whereas producing the $3<z<4$ positive gradients from observations would require substantial changes to the galactic winds and ISM of the models, at $z>6$ only a subtle change (if, indeed, one is needed) would be required to flatten simulated gradients to their observed counterparts.
This correction could come in the form of turbulent metal diffusion between gas elements within the ISM.
EAGLE and SIMBA -- owing to their SPH and MFM implementations, respectively -- do not have any metals advected from one element to the next.
Metals in these models are locked in the gas particle in which they form.
This has been shown to produce metal distributions that are more inhomogeneous than observed \citep[see, e.g.,][]{Aguirre_2005}.
For Illustris and TNG, the MVM implementation of {\sc arepo} naturally allows metals to advect as the cells deform and reshape.
Beyond the cell deformation, however, there is no implementation for the transport of metals from small, unresolved turbulent eddies that drive diffusion in the ISM \citep[as in, e.g.,][]{Smagorinsky_1963,Shen_2010,Semenov_2016,Su_2017,Escala_2018,Semenov_2024}.
The addition of these turbulent eddies in these models would have the effect of redistributing metals through the ISM, potentially flattening the metallicity gradient in the process.
\cite{Bellardini_2021} show this concretely using the FIRE model and a range of metal diffusion coefficients.
\cite{Bellardini_2021} find that the lack of a diffusion coefficient significantly steepens a gradient while higher coefficients allow for flattening.
It is thus possible that the subtle differences between the metallicity gradients at these high redshifts indicate that these subgrid models require a metal diffusion model to redistribute the metals of the ISM.

In summary, it is as of yet uncertain whether the tension between the simulations and observations at $z>6$ is meaningful.
Current and upcoming observational campaigns with JWST and ALMA should help clarify the picture by providing additional galaxies for more robust comparisons.

\section{Conclusions}
\label{sec:conclusion}

In this work, we analyze the gas-phase metallicity gradients of star forming galaxies across a wide mass ($10^{8.0}M_\odot < M_\star \lesssim 10^{12.0}M_\odot$) and redshift range ($0\leq z\leq8$) in EAGLE, Illustris, IllustrisTNG, and SIMBA.
We construct face-on metallicity maps and reduce them into a metallicity radial profile that we fit with a single linear regression in the ``star forming region'' of the galaxies (see Figure~\ref{fig:methods_fig}).

Our conclusions are as follows:
\begin{itemize}[leftmargin=10pt]
    \item We find that the evolution of metallicity gradients in EAGLE, Illustris, TNG, and SIMBA are all very similar, with more negative gradients further back in time (Section~\ref{subsec:each_sim_gradients} and Figure~\ref{fig:summary_fig}).
    We speculate that this is likely owing to the relatively smooth implementation of stellar feedback that arises naturally from the subgrid ISM prescription common amongst the models analyzed in this work.
    
    \item In more detail, we find that different stellar mass bins have different metallicity gradient evolution (Section~\ref{subsection:evolution_by_mass} and Figure~\ref{fig:evolution}).
    We find that galaxy stellar masses of $10^{8.0}M_\odot\,-\,10^{9.0}M_\odot$ have virtually no redshift evolution out to $z=8$ (although gradients are slightly more negative around cosmic noon).
    On the other hand, galaxies with stellar masses of $10^{10.0}M_\odot-10^{11.0}M_\odot$ have the strongest redshift evolution.
    Meanwhile, galaxies of lower intermediate ($10^{9.0}M_\odot\,-\,10^{10.0}M_\odot$) and very high masses ($10^{11.0}M_\odot\,-\,~10^{12.0}M_\odot$) have a moderate amount of redshift evolution.
    

    \item We find that much of the mass-dependent redshift evolution of the gradients disappears when normalized by galaxy size (Figure~\ref{fig:normalized_by_size}).
    In this scenario, galaxies build-up their metallicity gradient quickly in the early Universe, \edit{which} then persist relatively uninterrupted owing to the relatively smooth feedback produced by the equation of state ISM.
    The metallicity gradient only evolves to flatter ${\rm dex/kpc}$ values at lower redshift as the galaxies grow in physical size.

    \item Finally, we compare the simulation results to those of observations (Section~\ref{subsec:comparison_with_obs}).
    We find that our results are in contrast with observations at low-to-intermediate redshifts ($1<z<4$) which exhibit a larger fraction of positive and flat gradients (see Section~\ref{subsubsec:3<z<4}).
    Comparing with higher redshift ($z>6$) JWST and ALMA observations, however, the tension is more subtle: observed gradients at $z>6$ are only slightly flatter than the simulations (see Section~\ref{subsubsec:z>6})..
    These comparisons suggest that metals may be under-mixed in many widely-used ISM subgrid models.
\end{itemize}

The spatial distribution of metals within the ISM of galaxies is critically sensitive to the underlying physics, particularly the feedback.
The upcoming prospects for high-redshift observations of metallicity gradients with, e.g., JWST provide an exciting opportunity to understand the processes driving galactic evolution as well as constrain future simulation models of galaxies.
A particular opportunity presents itself: no galaxy metallicity gradients have been observed in the redshift range of $4<z<6$ to date.
This redshift range is where the smooth feedback simulated galaxies ubiquitously have strong negative metallicity gradients, regardless of stellar mass.
Filling in this observational gap would therefore provide key insights into the level of metal mixing required in the ISM of simulated galaxies.

\section*{Data Availability}

Raw data from the EAGLE, Illustris, TNG, and SIMBA simulations is publicly available. EAGLE: \href{https://icc.dur.ac.uk/Eagle/database.php}{https://icc.dur.ac.uk/Eagle/database.php},
Illustris: \href{https://www.illustris-project.org/data/}{https://www.illustris-project.org/data/}, IllustrisTNG: \href{https://www.tng-project.org/data/}{https://www.tng-project.org/data/}, and SIMBA \href{http://simba.roe.ac.uk/}{http://simba.roe.ac.uk/}.
\edit{We provide the simulation data presented in each of Figure~\ref{fig:summary_fig}-\ref{fig:Rsfr_evo} at \href{https://github.com/AlexGarcia623/gradients_paper1_data}{github.com/AlexGarcia623/gradients\_paper1\_data}.
The original scripts used to generate the complete data footprint (and all other scripts used to support the findings of this work) are publicly available here: \href{https://github.com/AlexGarcia623/gradients_paper1_scripts}{github.com/AlexGarcia623/gradients\_paper1\_scripts}.
}

\begin{acknowledgments}
\edit{We thank the anonymous referee for their thoughtful comments that helped improve the quality of this manuscript.}
We thank Benedetta Ciardi for helpful conversations that led to the success of this paper.
AMG thanks Carol Garcia for her help in the design of several of the figures in this work.

The authors acknowledge Research Computing at The University of Virginia, University of Florida Information Technology, and Max Planck Computing and Data Facility for providing computational resources and technical support that have contributed to the results reported within this publication.
We acknowledge the Virgo Consortium for making their simulation data available. The EAGLE simulations were performed using the DiRAC-2 facility at Durham, managed by the ICC, and the PRACE facility Curie based in France at TGCC, CEA, Bruy\`eresle-Ch\^atel.

AMG acknowledges support from a Virginia Space Grant Consortium Graduate STEM Research Fellowship.
AMG and PT acknowledge support from NSF-AST 2346977 and the NSF-Simons AI Institute for Cosmic Origins which is supported by the National Science Foundation under Cooperative Agreement 2421782 and the Simons Foundation award MPS-AI-00010515.
KG is supported by the Australian Research Council through the Discovery Early Career Researcher Award (DECRA) Fellowship (project number DE220100766) funded by the Australian Government. 
KG is supported by the Australian Research Council Centre of Excellence for All Sky Astrophysics in 3 Dimensions (ASTRO~3D), through project number CE170100013. 
\end{acknowledgments}

%






\appendix

\section{On Simulation Mass Resolution Dependence}
\label{appendix:resolution}

\begin{figure*}
    \centering
    \includegraphics[width=\linewidth]{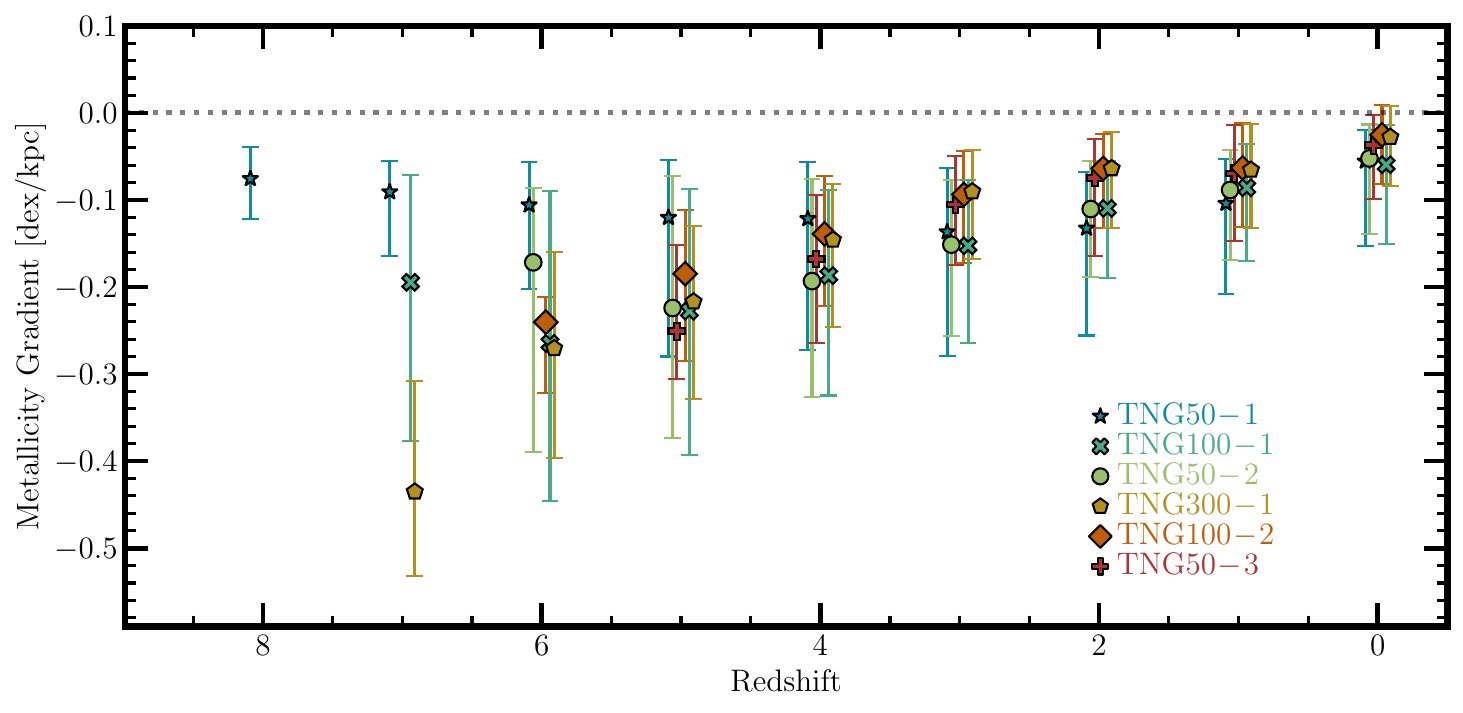}
    \caption{{\bf Comparison of several TNG resolution models.}
    Same as top panel of Figure~\ref{fig:summary_fig}, but for TNG50-1 (stars), TNG100-1 (Xs), TNG50-2 (circles), TNG300-1 (pentagons), TNG100-2 (diamonds), and TNG50-3 (plus) as a study of the impact of simulation resolution.
    Broadly speaking, TNG100-1 and TNG50-2 are comparable mass resolution to EAGLE ({\sc refL0100N1504}, as  analyzed in this work) and Illustris (Illustris-1, as analyzed in this work), whereas TNG300-1, TNG100-2, and TNG50-3 are comparable mass resolution to SIMBA (m50n512, as analyzed in this work).
    These data are presented in Table~\ref{tab:appendix_properties} (except for TNG50-1, which is in Table~\ref{tab:properties_1}).
    }
    \label{fig:appendix_summary}
\end{figure*}

The simulations analyzed in this work are of varying mass resolutions (see Section~\ref{subsec:simulations}).
We therefore dedicate this appendix to understanding the extent to which the results of this paper are dependent on the mass resolution of the chosen simulation.
The natural testing ground for this is the IllustrisTNG simulation suite as it has a (very) wide range of box sizes and mass resolutions within a fixed galaxy formation model.
We note that we do not modify the methodology of the main text in {\it any way} in this section (including the selection criteria based on the mass resolution of the simulation; see Sections~\ref{subsec:selection_criteria}~and~\ref{subsec:gradient_definitions}).

\begin{table*}
    \centering
    \begin{tabular}{crrrrrrrrrr}
\toprule
& \multicolumn{2}{c}{TNG50-2} & \multicolumn{2}{c}{TNG50-3} & \multicolumn{2}{c}{TNG100-1} & \multicolumn{2}{c}{TNG100-2} & \multicolumn{2}{c}{TNG300-1}\\\midrule
$m_{\rm baryon}~[M_\odot]$ & \multicolumn{2}{|c}{$3.6\times10^6$} & \multicolumn{2}{|c}{$2.9\times10^7$} & \multicolumn{2}{|c}{$7.5\times10^6$} & \multicolumn{2}{|c}{$6.0\times10^7$} & \multicolumn{2}{|c}{$5.9\times10^7$}\\
$L_{\rm box}~[{\rm Mpc}/h]$ & \multicolumn{2}{|c}{$35$} & \multicolumn{2}{|c}{$35$} & \multicolumn{2}{|c}{$75$} & \multicolumn{2}{|c}{$75$} & \multicolumn{2}{|c}{$205$}\\
\midrule
& \multicolumn{1}{c}{$N$} & $\nabla$ [dex/kpc] & \multicolumn{1}{c}{$N$} & $\nabla$ [dex/kpc] & \multicolumn{1}{c}{$N$} & $\nabla$ [dex/kpc] & \multicolumn{1}{c}{$N$} & $\nabla$ [dex/kpc] & \multicolumn{1}{c}{$N$} & $\nabla$ [dex/kpc] \\\midrule
$z=0$ & \lc{$1152$} & $-0.052^{-0.013}_{-0.139}$ & \lc{$226$} & $-0.037^{-0.003}_{-0.099}$ & \lc{$9281$} & $-0.060^{-0.014}_{-0.150}$ & \lc{$1573$} & $-0.025^{+0.009}_{-0.082}$ & \lc{$34510$} & $-0.028^{+0.008}_{-0.084}$ \\
$z=1$ & \lc{$1278$} & $-0.089^{-0.043}_{-0.169}$ & \lc{$289$} & $-0.070^{-0.014}_{-0.147}$ & \lc{$11312$} & $-0.086^{-0.036}_{-0.170}$ & \lc{$2363$} & $-0.063^{-0.012}_{-0.131}$ & \lc{$47948$} & $-0.066^{-0.012}_{-0.132}$ \\
$z=2$ & \lc{$954$} & $-0.110^{-0.056}_{-0.188}$ & \lc{$200$} & $-0.075^{-0.030}_{-0.164}$ & \lc{$8882$} & $-0.110^{-0.055}_{-0.189}$ & \lc{$1723$} & $-0.064^{-0.024}_{-0.133}$ & \lc{$34420$} & $-0.064^{-0.022}_{-0.132}$ \\
$z=3$ & \lc{$536$} & $-0.151^{-0.077}_{-0.257}$ & \lc{$74$} & $-0.105^{-0.049}_{-0.175}$ & \lc{$5280$} & $-0.153^{-0.077}_{-0.264}$ & \lc{$846$} & $-0.094^{-0.044}_{-0.173}$ & \lc{$15809$} & $-0.090^{-0.042}_{-0.168}$ \\
$z=4$ & \lc{$243$} & $-0.193^{-0.076}_{-0.327}$ & \lc{$28$} & $-0.168^{-0.094}_{-0.264}$ & \lc{$2501$} & $-0.187^{-0.089}_{-0.325}$ & \lc{$272$} & $-0.139^{-0.073}_{-0.222}$ & \lc{$5488$} & $-0.146^{-0.082}_{-0.246}$ \\
$z=5$ & \lc{$80$} & $-0.224^{-0.073}_{-0.373}$ & \lc{$10$} & $-0.251^{-0.152}_{-0.305}$ & \lc{$914$} & $-0.228^{-0.088}_{-0.393}$ & \lc{$72$} & $-0.185^{-0.112}_{-0.284}$ & \lc{$1450$} & $-0.217^{-0.130}_{-0.329}$ \\
$z=6$ & \lc{$25$} & $-0.172^{-0.086}_{-0.389}$ & \lcTwo{} & \lc{$271$} & $-0.265^{-0.090}_{-0.445}$ & \lc{$12$} & $-0.240^{-0.212}_{-0.322}$ & \lc{$232$} & $-0.270^{-0.160}_{-0.397}$ \\
$z=7$ & \lcTwo{} & \lcTwo{} & \lc{$67$} & $-0.195^{-0.071}_{-0.377}$ & \lcTwo{} & \lc{$23$} & $-0.435^{-0.308}_{-0.532}$ \\
$z=8$ & \lcTwo{} & \lcTwo{} & \lcTwo{} & \lcTwo{} & \lcTwo{} \\\midrule
${\rm dex/kpc/}\Delta z$ & \lcTwo{$-0.025 \pm 0.005$} & \lcTwo{$-0.040 \pm 0.007$} & \lcTwo{$-0.027 \pm 0.005$} & \lcTwo{$-0.034 \pm 0.004$} & \lcTwo{$-0.052 \pm 0.008$}\\
\bottomrule
    \end{tabular}
    \caption{{\bf Average Gradients in the Varying Mass Resolution and Box Size TNG Samples.}
    The top two rows show the initial baryon mass resolution ($m_{\rm baryon}$) and simulation box size ($L_{\rm box}$), respectively, for each of the TNG runs analyzed in Appendix~\ref{appendix:resolution}.
    The remainder of the table is organized identically to that of Table~\ref{tab:properties_1}.
    }
    \label{tab:appendix_properties}
\end{table*}

We sample generously from the TNG suite (see Section~\ref{subsubsec:TNG}, and references therein, for a discussion of the TNG physics model).
We make use of three boxes from TNG50 suite of $(35~{\rm Mpc}/h)^3$ simulations: TNG50-1 (the same as the main body of this work) with $m_{\rm baryon}=8.5\times10^4M_\odot$, TNG50-2 with $m_{\rm baryon}=3.6\times10^6M_\odot$, TNG50-3 with $m_{\rm baryon}=2.9\times10^7M_\odot$. 
In addition, we use two boxes from the TNG100 suite of $(75~{\rm Mpc}/h)^3$ boxes (TNG100-1, $m_{\rm baryon}=7.5\times10^6M_\odot$; TNG100-2, $m_{\rm baryon}=6.0\times10^7M_\odot$) as well as one from the TNG300 suite of $(205~{\rm Mpc}/h)^3$ boxes (TNG300-1, $m_{\rm baryon}=5.9\times10^7M_\odot$).
The first advantage of using the TNG suite for this analysis is that TNG50-2 and TNG100-1 are comparable mass resolution to EAGLE ($m_{\rm baryon}=1.81\times10^6M_\odot$) and Illustris ($m_{\rm baryon}=1.26\times10^6M_\odot$), while TNG50-3, TNG100-2, and TNG300-1 are comparable to the mass resolution of SIMBA ($m_{\rm baryon}=1.28\times10^7$).
The second advantage is that the comparable resolutions span multiple box sizes and, consequently, varying mass ranges (see Section~\ref{subsec:selection_criteria}~and~\ref{subsection:evolution_by_mass}).

Figure~\ref{fig:appendix_summary} shows the redshift evolution of gradients in each TNG simulation (data also presented in Table~\ref{tab:appendix_properties}).
In general, we find the same qualitative trend as in Section~\ref{subsec:each_sim_gradients}: increasingly negative gradients at higher redshift.
The gradients become more negative at rate of $-0.025~{\rm dex/kpc/}\Delta z$ in TNG50-2, $-0.040~{\rm dex/kpc/}\Delta z$ in TNG50-3, $-0.027~{\rm dex/kpc/}\Delta z$ in TNG100-1, $-0.034~{\rm dex/kpc/}\Delta z$ in TNG100-2, and $-0.052~{\rm dex/kpc/}\Delta z$ in TNG300-1.
Interestingly, these represent moderately-to-significantly stronger evolution than in the main text, which are closer to $-0.015~{\rm dex/kpc/}\Delta z$ (with the exception of SIMBA).
Regardless, we note that the gradients are remarkably well converged in the TNG model, though some variance exists.

\begin{figure*}
    \centering
    \includegraphics[width=\linewidth]{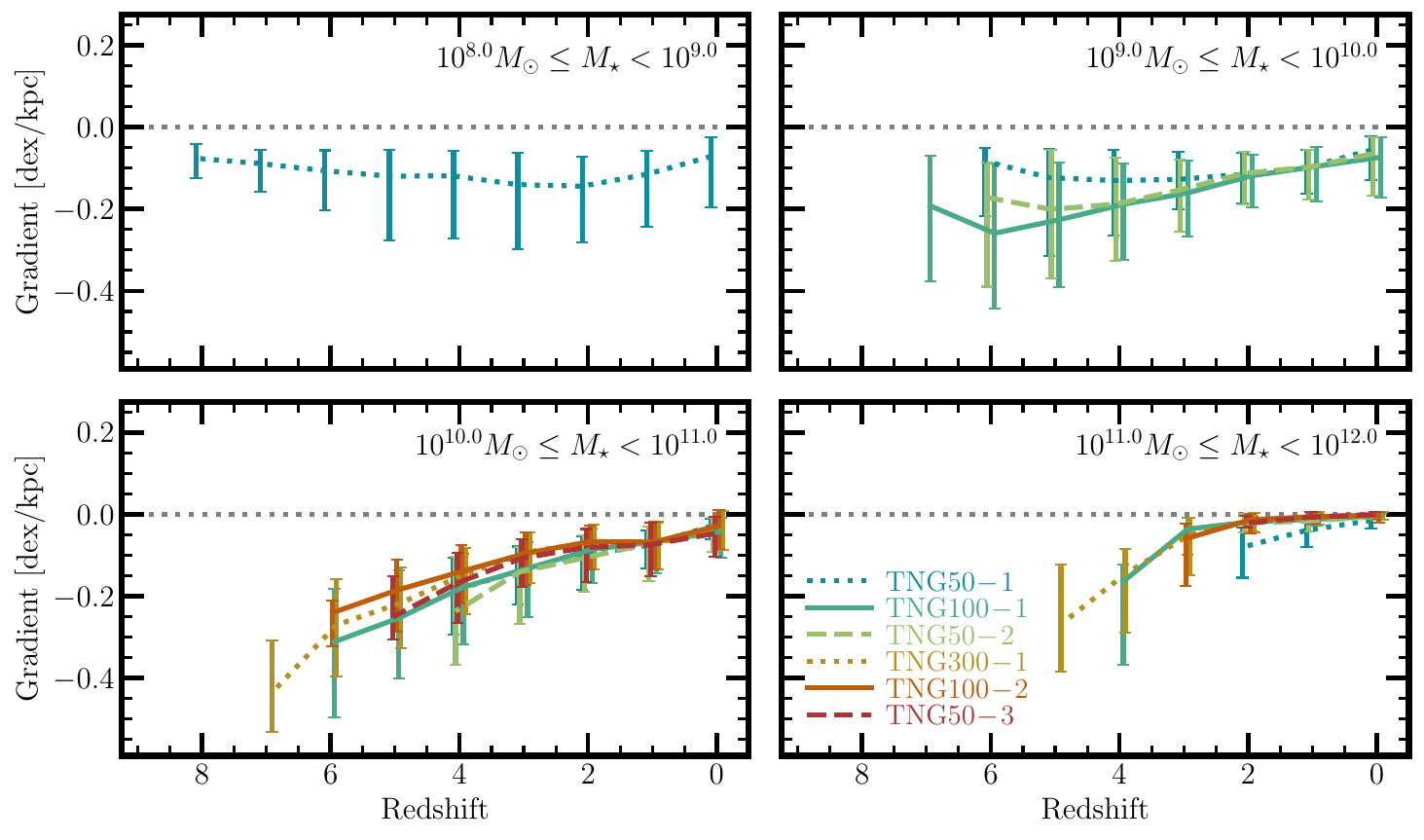}
    \caption{{\bf Metallicity Gradient Evolution in TNG Suite by Stellar Mass.}
    Same as Figure~\ref{fig:evolution} now with the six different TNG simulations.
    We note that the top left panel is identical to that of Figure~\ref{fig:evolution} as none of the other TNG simulations are high enough mass resolution to robustly measure gradients at $M_\star < 10^{9.0}M_\odot$.
    }
    \label{fig:appendix_by_mass}
\end{figure*}

We argue that this variance can again be explained by the different mass ranges of the simulations.
Figure~\ref{fig:appendix_by_mass} shows the evolution of the metallicity gradients broken down into the different stellar mass bins (similar to Figure~\ref{fig:evolution}).
The same general trend as noted in Section~\ref{subsection:evolution_by_mass} is true here: the strongest metallicity gradient evolution is present in galaxies with masses $10^{10.0}M_\odot\leq M_\star < 10^{11.0}M_\odot$, with weaker evolution at higher ($10^{11.0}M_\odot\leq M_\star < 10^{12.0}M_\odot$) and lower masses ($10^{9.0}M_\odot\leq M_\star < 10^{10.0}M_\odot$).
It should again be noted that these results are {\it remarkably} well converged.
Given that the physics model is identical between the different runs, the source of the differences likely come from two sources.
The first is cosmic variance; while the same box size runs (e.g., TNG50-1, 50-2, and 50-3) were run with the same initial conditions, the larger boxes have different initial conditions.
We should therefore expect slightly different populations of galaxies in each box size.
The second is that the gradient evolution mass relationship is {\it not} discrete, rather it is more continuous than the summary four stellar mass bins suggest.
This can be seen in the difference between TNG50-1 and 50-2 in the top right panel of Figure~\ref{fig:appendix_by_mass}.
At $z=0-2$ the gradients are virtually identical between 50-1 and 50-2, however, at $z>3$, they begin to diverge.
What is happening here is that the underlying sample of galaxies is very similar (and large) in TNG50-1 and 50-2 at $z=0-2$, but at higher redshift there are fewer galaxies overall.
Small changes in the populations, such as there being more low-mass galaxies in TNG50-1 and high-mass galaxies in TNG50-2 at $z>3$, have significant impacts on the resultant gradients.
This shifts the TNG50-1 distribution ``closer'' to that of the $M_\star < 10^9M_\odot$ galaxies and TNG50-2 ``closer'' to that of the $10^{10.0}M_\odot \leq M_\star < 10^{11.0} M_\odot$ causing the divergence we see.

It is also worth highlighting that in the highest mass bin, at $z>4$, the gradients have significant gradient evolution not seen in Figure~\ref{fig:evolution}.
Its absence in Figure~\ref{fig:evolution} is due to the smaller boxes of the main text.
It is therefore difficult to make meaningful statements about galaxies of that mass in these high redshifts.
Here, however, we find that there is some evidence that these massive galaxies have quite strong negative gradients at these early times.

In summary, we do not find that there is any significant mass resolution dependencies in the results of this work.

\section{Different Definitions of Gradients}
\label{appendix:gradient_definitions}

\begin{figure*}
    \centering
    \includegraphics[width=\linewidth]{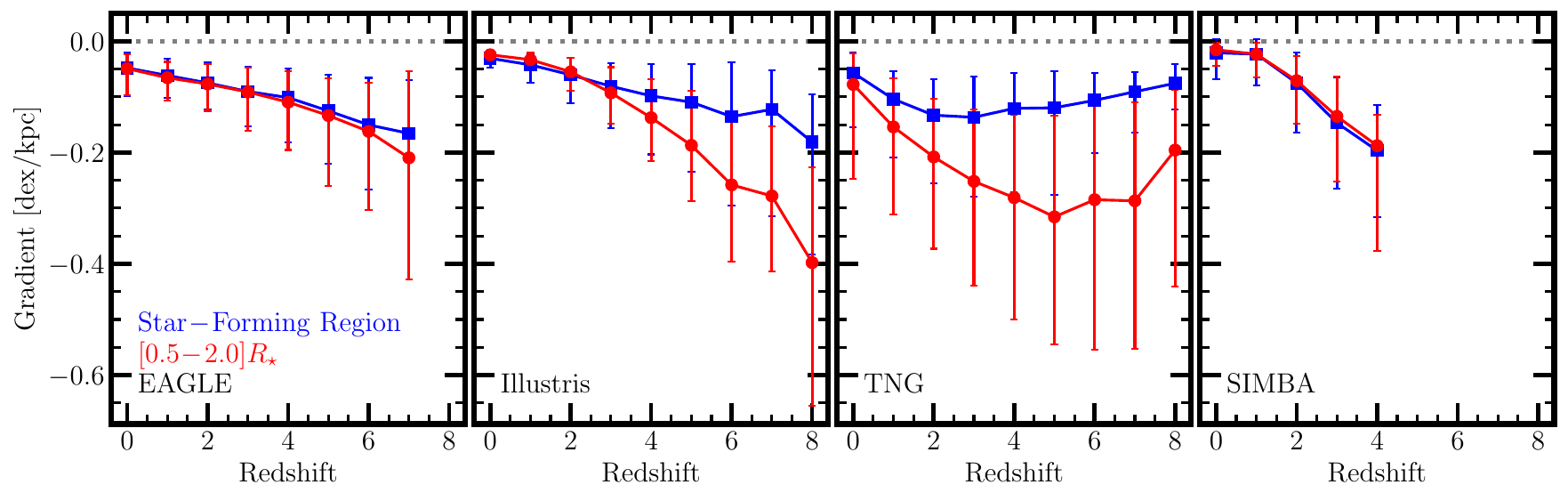}
    \caption{{\bf Comparison of Gradient Methodology.} Comparison of the star forming region gradient (as in the main body of this work; squares) and a fixed galactic size (as in, e.g., \protect\citeauthor{Sanchez_2012} \protect\citeyear{Sanchez_2012}, \protect\citeyear{Sanchez_2014}; \protect\citeauthor{Sanchez_Menguiano_2016} \protect\citeyear{Sanchez_Menguiano_2016}; \protect\citeauthor{Belfiore_2017} \protect\citeyear{Belfiore_2016}; circles) for the four different simulation models analyzes in this work.}
    \label{fig:appendix_method_comparison}
\end{figure*}

Our methodology of calculating the metallicity gradients is not the only way by which one could calculate a gradient.
Many low redshift observational works instead calculate the gradient within a fixed galactic size: from $0.5R_{\rm eff}$ to $2.0R_{\rm eff}$ \citep{Sanchez_2012,Sanchez_2014, Sanchez_Menguiano_2016,Belfiore_2017}.
We therefore refit the gradients with the fixed size criteria in this section (keeping all other methods the same) as a point of comparison.
We present this comparison in Figure~\ref{fig:appendix_method_comparison}.

We find that generally the population of gradients is nearly indistinguishable between the two methods in EAGLE and SIMBA.
In TNG, and $z\geq4$ in Illustris, however, there seems to be some disagreement between the two methods.
This behavior is the likely the result of two related effects.
Firstly, at $z\gtrsim4$ in Illustris the typical extent of the star forming region stretches beyond $2.0R_\star$.
In \cite{Garcia_2023}, we show that the ``break radius'' (the radius where the steep inner gradient begins to flatten out) is at $\sim2.5 R_\star$ for stacked galaxies fairly uniformly across redshift in Illustris.
The fixed size method is thus guaranteed never to encapsulate this transition; yet, once the star forming region method is more extended than $2.0R_\star$, on the other hand, it may begin to include this transition to a shallower gradient.
This would have the effect of flattening the overall derived gradient from the star forming region method, as is seen.
In TNG, $0.5R_\star$ tends to probe slightly {\it closer} to the galaxy center than $R_{\rm in}^\prime$.
We specifically choose not to fit these central parts of the gradients as they frequently deviate quite significantly from the expected linear fit.
In particular, they often are much steeper than the true gradient (as is the case within $<1$ kpc of the galaxy presented in Figure~\ref{fig:methods_fig}).
A steeper gradient is therefore more often obtained from this fixed size method since the fit encompasses these nearer-to-the-center regions with a steep local gradient.

It is interesting to note that with this method of deriving gradients there may be a more significant tension between the Illustris and TNG models and observations at high-redshift.
We suspect that the methodology of current high-redshift gradient measurements is more similar to the star-forming region analysis, however, as the recent high redshift observations do not make a specific cut on galaxy size \citep{Troncoso_2014,Wang_2022,Arribas_2024,Venturi_2024,Vallini_2024}.
It is therefore more likely that the observations are picking up on regions with bright emission lines from star-forming regions without regard to the size of the galaxy.

\section{Different Gradient Evolution Regression Technique}
\label{appendix:regression}

\begin{table*}
    \centering
    \begin{tabular}{llx{0.13\linewidth}x{0.13\linewidth}x{0.13\linewidth}x{0.13\linewidth}x{0.13\linewidth}}
        \toprule
        & & \multicolumn{5}{c}{Stellar Mass Bin $[\log M_\odot]$} \\\cmidrule(lr){3-7}
        Simulation & & All Combined & $8.0-9.0$ & $9.0-10.0$ & $10.0-11.0$ & $11.0-12.0$ \\\midrule
        EAGLE     & $[{\rm dex/kpc}/\Delta z]$ & $-0.018\pm0.000$ & -- & $-0.016\pm0.000$ & $-0.028\pm0.001$ & $0.004\pm0.008$ \\
                  & $[{\rm dex/}R_{\rm SFR}/\Delta z]$ & & -- & $-0.028\pm0.001$ & $-0.034\pm0.001$ & $0.002\pm0.005$ \\\midrule
        Illustris & $[{\rm dex/kpc}/\Delta z]$ & $-0.022\pm0.000$ & -- & $-0.022\pm0.000$ & $-0.023\pm0.000$ & $-0.013\pm0.001$ \\
                  & $[{\rm dex/}R_{\rm SFR}/\Delta z]$ & & -- & $-0.014\pm0.000$ & $-0.013\pm0.000$ & $-0.004\pm0.001$ \\\midrule
        TNG       & $[{\rm dex/kpc}/\Delta z]$ & $-0.011\pm0.000$ & $-0.004\pm0.001$ & $-0.018\pm0.001$ & $-0.041\pm0.002$ & $-0.033\pm0.004$ \\
                  & $[{\rm dex/}R_{\rm SFR}/\Delta z]$ & & $-0.005\pm0.001$ & $-0.018\pm0.001$ & $-0.051\pm0.003$ & $-0.020\pm0.005$ \\\midrule
        SIMBA     & $[{\rm dex/kpc}/\Delta z]$ & $-0.045\pm0.003$ & -- & -- & $-0.045\pm0.003$ & $-0.051\pm0.012$ \\
                  & $[{\rm dex/}R_{\rm SFR}/\Delta z]$ & & -- & -- & $-0.025\pm0.007$ & $-0.035\pm0.008$ \\
        \bottomrule
    \end{tabular}
    \caption{{\bf Comparison of Regression Methodology.}
    The best-fit linear regression parameters using all galaxies, opposed to Tables~\ref{tab:properties_1}~and~\ref{tab:evolution_by_mass} which use regressions through the median weighted by number of galaxies at each mass bin.
    }
    \label{tab:appendix_evolution_by_mass}
\end{table*}

Throughout this work we quote the evolution of metallicity gradients in terms of a linear regression to the median gradients as a function of redshift weighted by the number of galaxies in each mass bin.
Table~\ref{tab:appendix_evolution_by_mass} shows the results of a linear regression using all gradients instead.
We note that this method is more sensitive to individual galaxy outliers. 
This can be particularly clearly seen in the case of the $10^{10.0} M_\odot\leq M_\star < 10^{11.0}M_\odot$ bin, where (by visual inspection of Figure~\ref{fig:normalized_by_size}) there is no strong evolution of ${\rm dex/}R_{\rm SFR}$ gradients until $z\gtrsim4$ in each simulation.
Yet the (relatively) few galaxies that do have some strong ${\rm dex/}R_{\rm SFR}$ gradients at these early times (reasons for which are speculated on in Section~\ref{subsec:mass_dependence_discussion}) have a significant impact on the overall derived gradient evolution.
This is because our fitting methodology (\verb|numpy.polyfit|)\ignorespaces
\footnote{\ignorespaces
\href{https://numpy.org/doc/stable/reference/generated/numpy.polyfit.html}{https://numpy.org/doc/stable/reference/generated/numpy.polyfit.html}
}
uses a least squares algorithm, punishing significant outliers.

\section{Mass Dependence at Fixed Redshift}
\label{appendix:mass_evo_fixed_z}

\begin{figure}
    \centering
    \includegraphics[width=\linewidth]{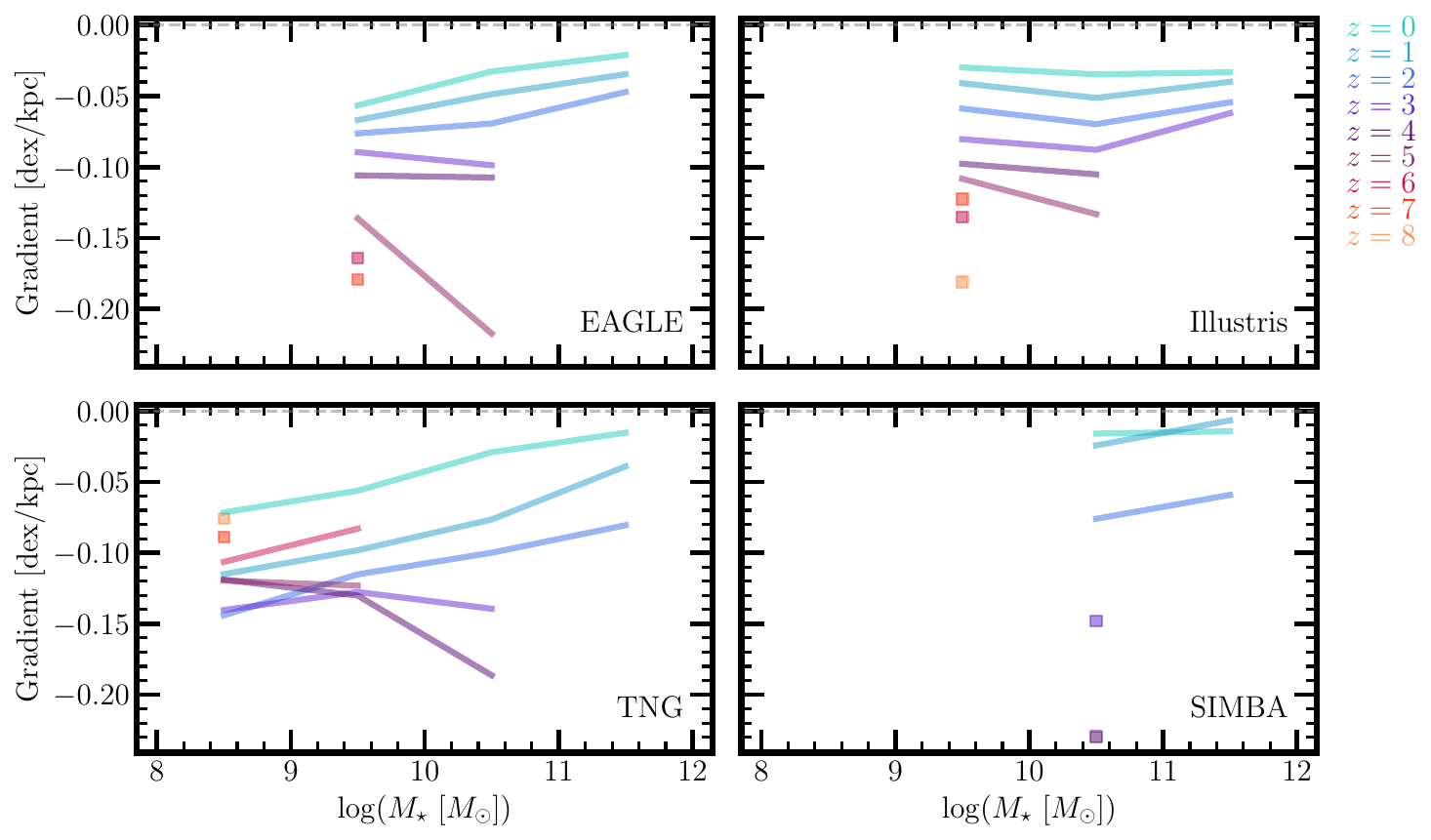}
    \caption{{\bf Stellar Mass Evolution of Gradients in Each Simulation.}
    Same as Figure~\ref{fig:evolution} except instead of redshift on the abscissa, we plot the gradients as a function of stellar mass for EAGLE (top left), Illustris (top right), TNG (bottom left), and SIMBA (bottom right).
    We plot the median gradient in stellar mass bins of 1.0 dex (as in the same bins of the four panels of Figure~\ref{fig:evolution}).
    Where there is only one mass bin available, we plot a single square in that mass bin.
    }
    \label{fig:appendix_D1}
\end{figure}

\edit{
Figure~\ref{fig:evolution} shows the evolution of metallicity gradients as a function of redshift broken into different stellar mass bins.
While it is possible to determine the mass trends at fixed redshift from these plots, it is not done straightforwardly.
We therefore provide Figure~\ref{fig:appendix_D1} showing the gradient evolution as a function of mass at fixed redshift for EAGLE (top left), Illustris (top right), TNG (bottom left) and SIMBA (bottom right).
We note that we use the same four mass bins as the four panels of Figure~\ref{fig:evolution} ($10^{8.0-9.0}M_\odot$, $10^{9.0-10.0}M_\odot$, $10^{10.0-11.0}M_\odot$, and $10^{11.0-12.0}M_\odot$) and place each bin at the middle of the mass bin.
For some simulations at some redshift, there is only one mass bin.
In these cases, we plot only a single square.
}

\edit{
We note that all of the data is unchanged with respect to Figure~\ref{fig:evolution}; however, we briefly discuss the mass trends here.
We find that there is generally a trend of flatter gradients with increasing stellar mass at low redshift ($z\lesssim2$).
At higher redshift ($z\sim3-5$), the trend appears to start to invert such that galaxies at higher mass have {\it stronger} gradients in EAGLE and Illustris (although TNG complicates this, see below).
The stronger gradients at larger masses in high redshift systems is broadly consistent with the interpretation of Figure~\ref{fig:evolution}: higher mass galaxies have stronger redshift evolution.
Therefore, at $z\sim3-5$, the strong negative gradients are comparable to, if not stronger than, their lower mass counterparts.
However, the picture is complicated by TNG.
We suspect that this may be due in part to the redshift scaling winds in TNG, which increase the efficiency of the feedback at high redshift \citep{Pillepich_2018a}.
The increased feedback would be able to provide additional metal mixing at these early times to flatten the gradient (compared to models without the redshift scaling winds).
At $z\geq6$, the mass trends are difficult to tell as there is one (if any) mass bin populated in each simulation.
}

\section{Size Evolution of Galaxies}
\label{appendix:size_evo}

\begin{figure}
    \centering
    \includegraphics[width=\linewidth]{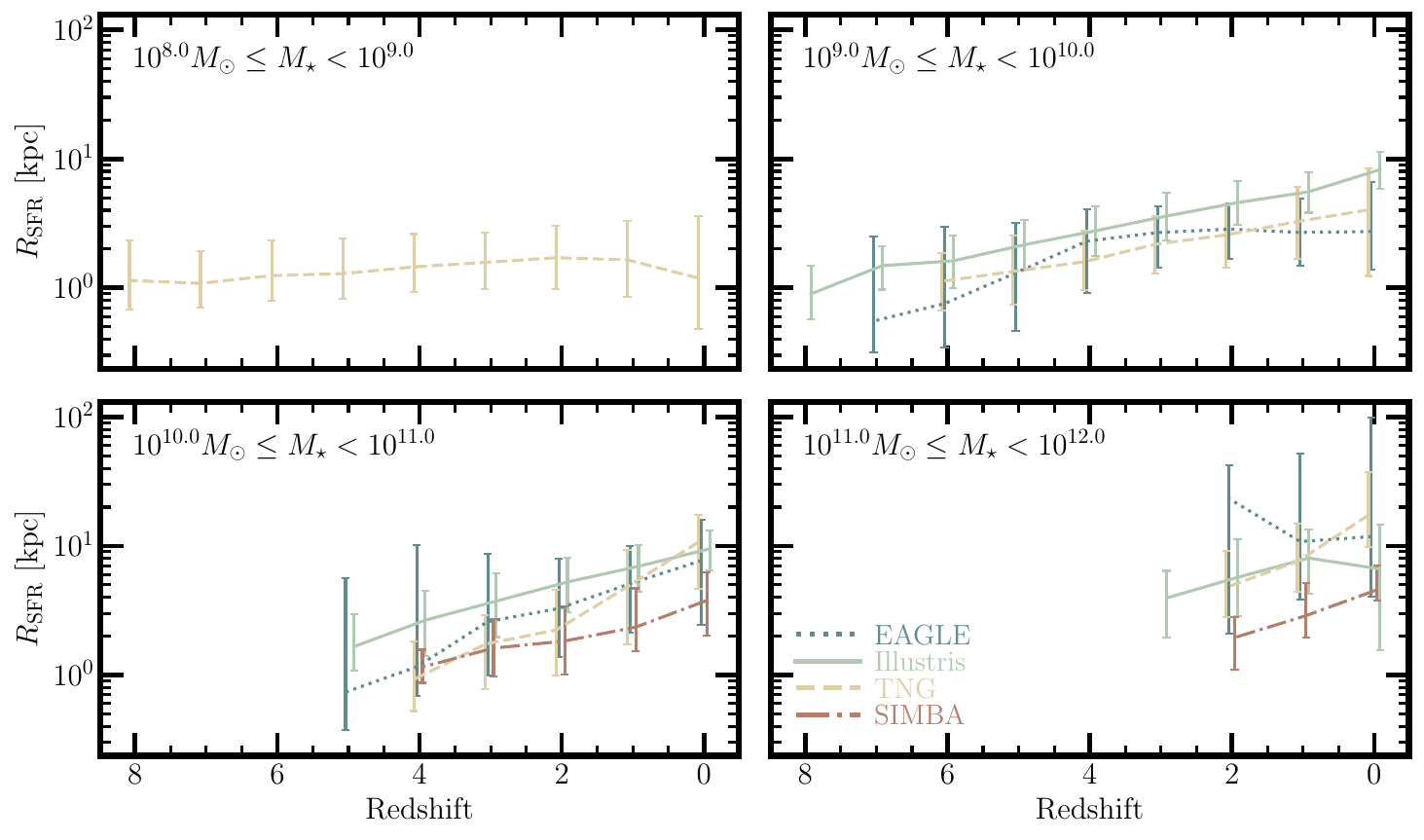}
    \caption{{\bf Evolution of Galaxy Size.}
    The size evolution ($R_{\rm SFR}$, the radius enclosing half of the star formation within a galaxy) of galaxies in EAGLE (dotted blue line), Illustris (solid green line), TNG (dashed yellow line), and SIMBA (dot dashed brown line) broken into four galaxy stellar mass bins (following the convention of Figures~\ref{fig:evolution}~and~\ref{fig:normalized_by_size}).
    }
    \label{fig:Rsfr_evo}
\end{figure}

\edit{Figure~\ref{fig:normalized_by_size} shows the evolution of metallicity gradients normalized by their size ($R_{\rm SFR}$) in each simulation.
In this appendix, we report and briefly comment on the sizes of the galaxies.
}

\edit{Figure~\ref{fig:Rsfr_evo} shows the $R_{\rm SFR}$ (the radius enclosing 50\% of the star formation in the galaxy) evolution of galaxies in each of the simulations, broken into the four different stellar mass bins.
We first comment that we required the star-forming region, over which we compute the gradient, to be $>1$ kpc (see Section~\ref{subsec:gradient_definitions}).
In Figure~\ref{fig:Rsfr_evo}, however, there are galaxies with sizes less than $1$ kpc, particularly those with low stellar masses and at the highest redshifts.
The reasoning for this discrepancy is that we do not make the size cut on $R_{\rm SFR}$, we do it on the star-forming region.
Therefore, the ``size'' of the galaxy (as we define here) can be $<1$ kpc, while the region over which we compute the gradient (i.e., the star-forming region) is $>1$ kpc.
Moreover, we note that the sample of galaxies at the lowest masses and highest redshifts is biased based on our star formation region size cut.
We preferentially omit the small, compact star formation regions of galaxies for resolution reasons (see Section~\ref{subsec:gradient_definitions} for more discussion on this point).
}

\edit{
We find that galaxies are typically smaller at higher redshift and lower stellar masses.
Both the stellar mass and redshift trends are qualitatively similar to those seen in observations \citep{vanderWel_2014,Ormerod_2024,Cook_2025}.
There are two notable exceptions to this: (i) the lowest mass bin in TNG and (ii) the highest mass bin of EAGLE.
In TNG, galaxies with stellar mass $10^{8.0}M_\odot\leq M_\star < 10^{9.0}M_\odot$ have sizes that are generally a few kpc at $z=0$, which raises slightly to $z=1$ and then decreases gradually further back in time.
The variation between redshifts here is quite subtle, but is likely related to the redshift scaling winds in TNG, which work to suppress star formation at low redshift \citep{Pillepich_2019}.
This suppression of star formation likely preferentially causes the relatively minor decrease in $R_{\rm SFR}$ for the smallest (i.e., least massive) TNG galaxies.
In EAGLE, galaxies with stellar mass $10^{11.0}M_\odot\leq M_\star < 10^{12.0}M_\odot$ are highly extended with $R_{\rm SFR}$ of $\gtrsim10$ kpc at $z=0-2$.
Of particular interest here is that the size of the galaxies actually increases to $z=2$.
We note that this behavior is not seen in other proxies for the size of the galaxies (e.g., the stellar half mass radius).
The extended star formation at $z=2$ in the most massive EAGLE galaxies may driven by a combination of high baryon fractions in the CGM, recycled outflows, and lack of preventative inflows caused by AGN feedback \citep{Wright_2024}.
These three ingredients provide fuel for star formation at larger radii in EAGLE and preferentially prevent inflows in the ISM.
The star formation would therefore be expected to be more extended.
}

\edit{
We now briefly compare the overall normalization of sizes of galaxies and note that a detailed examination of the difference in the sizes of galaxies (and how that changes with redshift) is its own separate investigation.
In the intermediate mass bins ($10^{9.0-10.0}M_\odot$ and $10^{10.0-11.0}$) galaxies in Illustris tend to have the largest sizes, with EAGLE, TNG, and SIMBA having relatively similar sizes.
The large galaxies is one was of the main deficiencies of the original Illustris model \citep{Genel_2014}.
At the highest masses ($10^{11.0-12.0}M_\odot$), however, EAGLE galaxies have the most extened star formation (see above), Illustris and TNG have very similar sizes, and SIMBA is the most compact.
}


\bibliography{paper}{}
\bibliographystyle{aasjournal}



\end{document}